\documentclass[reprint, nofootinbib, amsmath,amssymb, aps, prd
]{revtex4-2}
\usepackage[utf8]{inputenc}
\usepackage{graphicx}
\usepackage{dcolumn}
\usepackage{xcolor}
\usepackage{bm}
\usepackage{hyperref}
\usepackage{orcidlink}
\usepackage{acronym}
\usepackage{xspace}

\makeatletter
\newcommand{\raisemath}[1]{\mathpalette{\raisem@th{#1}}}
\newcommand{\raisem@th}[3]{\raisebox{#1}{$#2#3$}}
\makeatother
\NewDocumentCommand{\newbbar}{O{0pt} O{0pt}}{
  \ensuremath{\mathrlap{\raisemath{#2}{\hspace*{#1}{\mathchar'26\mkern-9mu}}}b}%
}

\newcommand{\msun}{\ensuremath{M_\odot}}
\newcommand{\cheff}{\ensuremath{\chi_\mathrm{eff} }\xspace}
\newcommand{\D}{\ensuremath{\textnormal{d}}}

\begin{document}

\title{Seeking Spinning Subpopulations of Black Hole Binaries via Iterative Density Estimation}

\author{Jam Sadiq\orcidlink{0000-0001-5931-3624}\textsuperscript{1, 2}}\thanks{Corresponding email: jamlucky.qau@gmail.com}
\author{Thomas Dent\orcidlink{0000-0003-1354-7809}\textsuperscript{3}}
\author{Ana Lorenzo-Medina\orcidlink{0009-0006-0860-5700}\textsuperscript{3}}

\affiliation{\textsuperscript{1} SISSA, Via Bonomea 265, 34136 Trieste, Italy and INFN Sezione di Trieste}
\affiliation{\textsuperscript{2} IFPU - Institute for Fundamental Physics of the Universe, Via Beirut 2, 34014 Trieste, Italy}
\affiliation{\textsuperscript{3} IGFAE, University of Santiago de Compostela, E-15782 Spain}

\date{\today}

\begin{abstract} 
  Attempts to understand the formation of binary black hole (BBH) systems detected via gravitational wave (GW) emission are affected by many unknowns and uncertainties, from both the observational and theoretical (astrophysical modelling) sides. 
  Binary component spins have been proposed as a means to investigate formation channels, 
  however obtaining clear inferences  
  is challenging, given the apparently low magnitude of almost all merging BH spins and their high measurement uncertainties.  Even for the effective aligned spin $\cheff$ which is more precisely measured than component spins, specific model assumptions have been required to identify any clear trends. 
  Here, we reconstruct the joint component mass and \cheff distribution of BBH mergers with minimal assumptions using the GWTC-3 catalog, using 
  an iterative kernel density estimation (KDE)-based method.  We reproduce some features seen in previous analyses, for instance a small but preferentially positive $\cheff$  
  for low-mass mergers; we also identify a possible subpopulation of higher-spin BBH with $|\cheff|$ up to $\sim\! 0.75$ for primary masses $m_1 \gtrsim 40\,\msun$, in addition to the bulk of the distribution with $|\cheff| \lesssim 0.2$.  This finding is consistent with previous studies indicating a broader spin distribution at high mass, suggesting a distinct origin for the high-spin systems.  We also identify a new potential trend of low-mass BBHs: the \emph{derivative} of $\cheff$ with respect to $m_1$ ($m_2$) is positive (negative) over the $10$--$15\,\msun$ range.  This apparent structure may be related to a previously reported anticorrelation between mass ratio and \cheff.  
\end{abstract}

\maketitle

\section{Introduction}
Ever since the first gravitational wave (GW) detection revealed a binary black hole (BBH) source with the---previously unsuspected---component masses of around 35\,\msun\ \cite{LIGOScientific:2016aoc,LIGOScientific:2016vpg}, 
LIGO-Virgo observations~\cite{LIGOScientific:2014pky,VIRGO:2014yos}  of compact binaries have continued to yield surprises. 
Having a set of detected compact binaries, such as the order(100) candidates catalogued in the GWTC releases~\cite{LIGOScientific:2018mvr,LIGOScientific:2020ibl,LIGOScientific:2021usb,KAGRA:2021vkt}, it is possible to study population properties of these compact binaries and eventually draw implications from these properties on binary astrophysical formation and evolution~\cite{LIGOScientific:2020kqk,KAGRA:2021duu}.  

Black hole spins have been proposed as a key diagnostic in understanding binary formation channels~\cite{Vitale:2015tea,Rodriguez:2016vmx,Farr:2017uvj,Farr:2017gtv}, though presenting persistent challenges in interpretation: recent studies towards this include~\cite{Bavera:2020inc,Zevin:2020gbd,Zevin:2022wrw,Broekgaarden:2022nst,Fuller:2022ysb,Payne:2024ywe,Banerjee:2024wbq, Antonini:2024het}. 
In isolated binaries, spins are expected to exhibit a preferred alignment with the orbital axis, breaking symmetry between prograde and retrograde configurations---assuming that magnitudes are not negligible~\cite[though see][]{Fuller:2019sxi}.  In contrast, dynamically formed
binaries in clusters should have randomized spin orientations~\cite{Sigurdsson1993,PortegiesZwart:1999nm,Rodriguez:2016kxx,Mapelli:2020vfa},
restoring approximate symmetry between aligned and in-plane spins
(with possible exceptions, e.g.\ triple systems~\cite {Silsbee:2016djf,Antonini:2017tgo}, post-encounter
tidal locking~\cite{Qin:2018vaa,Ma:2023nrf}
and AGN environments~\cite{Stone:2016wzz,Yang:2019cbr}). Hierarchical
mergers introduce distinct signatures: remnants from near-equal-mass
mergers tend to have dimensionless spin magnitudes $|\vec{s}|\sim 0.7$~\cite{Tichy:2008du,Gerosa:2017kvu,Tagawa:2021ofj},
and if retained in dense environments their orientations in subsequent mergers should again randomize.

However, observational limitations severely complicate such tests and signatures.  Current gravitational-wave observations alone are insufficient to draw definite conclusions about BBH formation channels or environments.  Most black hole spins are small or consistent with zero, in-plane components are largely unmeasurable for binary black holes, and even aligned spins often suffer from large uncertainties (which are correlated with binary mass ratio $q \equiv m_2/m_1$~\cite{Cutler:1994ys,Baird:2012cu,Ng:2018neg}). 
On the astrophysics side, the likelihood that multiple formation channels are contributing to the observed population~\cite{Zevin:2020gbd,Cheng:2023ddt} implies that first, inferences on the spin properties of detected BBH \emph{as a whole} are probably not a good representation of any one channel.  Therefore, direct interpretation of the inferred spin distribution without allowing for multiple subpopulations~\cite{Cheng:2023ddt} is likely misleading.  Secondly, as different channels may well be (partly or completely) separated over various dimensions of the BBH parameter space, i.e.\ component mass and possibly redshift~\cite[see e.g.][]{vanSon:2021zpk,Bavera:2022mef}, it will be critical to allow for \emph{correlations between spins and other BBH parameters}, including a possible mixture of \emph{subpopulations} which may mimic correlated trends. 

Given this complexity, the utility of spins remains limited: to make progress on the observational side, significant improvements in measurement precision (implying higher S/N) and a much greater sample size may be needed, and on the modelling side, more detailed and realistic scenarios encompassing multiple channels. 
To guide such efforts, it is crucial to make full use of currently available observations, including evidence for correlations or subpopulations: by refining empirical constraints on mass, spin, and redshift distributions, we can lay the groundwork for future studies to test formation models as the catalog grows.

Currently, some features in the BBH joint spin/mass/redshift distribution have been identified under various more or less restrictive model assumptions, and proposed as pointers towards formation channels. 
These assumptions or methods may be divided into parameterized models, where Bayesian hierarchical inference is used to determine posteriors for a small number of hyper-parameters corresponding to a relatively simple distribution function~\cite[e.g.][]{Mandel:2009pc,Thrane:2018qnx}; and semi- or non-parametric models, which aim to learn population properties from the data, either without requiring any specific functional form, or (for semi-parametric models) allowing for generalised deviations from a given parametric model~\cite{Powell:2019nmw, Tiwari:2020otp,Tiwari:2020vym,Tiwari:2021yvr,Rinaldi:2021bhm,Sadiq:2021fin,Edelman:2022ydv,Callister:2023tgi,Sadiq:2023zee,Toubiana:2023egi,Heinzel:2024jlc,Sadiq:2025aog}. 
Without attempting a complete summary of these findings, we can identify critical aspects of possible correlations between BH spins and other population properties. 

As mentioned earlier, many studies consider primarily the effective orbit-aligned spin $\cheff \equiv (s_{1z}+qs_{2z})/(1+q)$, which is more precisely measured than other spin variables~\cite{Ajith:2009bn} and so is often considered as a proxy for spin magnitude (though its sign is also informative on alignment). 
\cite{Tiwari:2021yvr} found via a flexible Gaussian mixture model that the magnitude $|s_z|$ (assumed identically distributed over binary components) has a wider distribution for chirp mass $\mathcal{M} \equiv (m_1m_2)^{3/5}/(m_1+m_2)^{1/5} \gtrsim 35\,\msun$. 
\cite{Callister:2021fpo} found an apparently significant negative correlation of $\cheff$ with $q$, though without attempting to identify its origin in BBH subpopulations or formation channels. 
\cite{Biscoveanu:2022qac,Bavera:2022mef} determined that the $\cheff$ distribution of widens at higher redshift, though such a trend is hard to distinguish from a correlation with mass given the strong selection effects that relate mass and distance~\cite{Fishbach:2018edt,Lorenzo-Medina:2024opt}.

More recently, efforts have been made to identify specific subpopulations with different spin properties~\cite[e.g.][]{Franciolini:2022iaa,Baibhav:2022qxm}. \cite{Godfrey:2023oxb} tailored a non-parametric mixture model taking as a reference the global peak of the local BBH merger distribution, with component masses $\sim\!10\,\msun$ and small positive $\cheff$, finding mild evidence for a high-mass ($m_1\gtrsim 50\,\msun$) subpopulation with a broader $\cheff$ distribution symmetric about zero.  By contrast, \cite{Ray:2024hos} found with a binned non-parametric analysis that the $30-40\,\msun$ component mass range containing the bulk of detections has a distinct $\cheff$ distribution symmetric about 0, suggesting a possible globular cluster origin~\cite{Antonini:2022vib}, while the remainder of the population tends to positive $\langle \cheff\rangle$.  Recently, \cite{Antonini:2024het} designed a change-point analysis to find a transition between a narrow, positive $\cheff$ distribution below, vs.\ a broad, uniform and symmetric $\cheff$ distribution above, a critical mass determined to be $\simeq 45\,\msun$, interpreting the higher-mass population as due to hierarchical formation in dense clusters.  Such an interpretation would be consistent with 1st generation BH (those formed from stellar collapse) having low spins and masses limited by pair-instability supernova (PISN) dynamics~\cite[for which see e.g.][]{Woosley:2021xba,vanSon:2021zpk}. 

Investigations using parametrized models of spin magnitude observables $|\vec{s}_{1,2}|$ have yielded comparable findings of a high-mass subpopulation with larger spins~\cite{Li:2023yyt,Li:2024jzi,Pierra:2024fbl}, though the higher number of degrees of freedom, including tilt angles, in this case may mean that detailed population properties are more affected by potential mismodeling bias. 

The lack of unanimity in findings when analyzing the same (GWTC-3~\cite{KAGRA:2021vkt}) data set reflects the high uncertainties in spin measurements (even if considering $\cheff$): due to these uncertainties the choice of model assumptions can strongly affect inferences on the spin distribution, and thus also the interpretation of individual events.  For parameterized models, it is generally not possible to quantify biases due to the model differing from the true population (mis-modelling) beyond self-consistency checks which, for data with large uncertainties, provide weak constraints.  Even for non-parametric methods, any assumption that the population distribution factorizes over different parameters (implying a lack of correlations) could lead to significant biases.  In particular, due to correlated measurement uncertainties in $\cheff$ and $q$, any assumption that the BBH mass, or mass ratio distribution takes a specific form may induce biases in spin inference (and vice versa). 

Therefore, to obtain an estimate of the joint spin-mass distribution we are led to consider, at a minimum, a fully three-dimensional non-parametric framework allowing for correlations between both BBH component masses and $\cheff$.  So far this approach  was only taken by~\cite{Ray:2024hos} using a binned (stepwise constant) model, where some correlations or features may be obscured by the (relatively large) bin size, and without reporting the full mass-spin distribution.  We demonstrated the technical feasibility of a 3d population reconstruction using a flexible adaptive KDE~\cite{Sadiq:2021fin,Sadiq:2023zee} in~\cite{Sadiq:2025aog}, where the ``resolution'' of the population estimate (i.e.\ how rapidly it may vary over BBH parameter space) is regulated by the local event density.\footnote{\cite{Heinzel:2024hva}, using a general finely-binned 2-d non-parametric method to investigate correlations, is the closest to us in approach and finds partly comparable results while restricted to the 2-d space of primary mass and $\cheff$.}  

In this work we then apply the same fully non-parametric, i.e.\ data-determined, method to reconstruct the BBH spin-mass distribution. 
The remainder of the paper is organized as follows: in section 2 we motivate and explain the method in outline, highlighting aspects where the application to effective spins differs from previous cases. 
In section 3 we apply our method to detected BBH in GWTC-3 using the available LIGO-Virgo-KAGRA (LVK) parameter estimation (PE)~\cite{ligo_scientific_collaboration_and_virgo_2021_5546663,ligo_scientific_collaboration_and_virgo_2021_5117970}; we compare the resulting mass distribution with our previous studies~\cite{Sadiq:2023zee} and assess the evidence for possible spin correlations or subpopulations.  In section 4 we discuss the implications of our results and consider possible further investigations.

\section{Method}

\subsection{Summary of multidimensional iterative adaptive KDE}

For our non-parametric population reconstruction, we employ a recently developed generalized iterative KDE method~\cite{Sadiq:2025aog} which extends our previous work~\cite{Sadiq:2021fin, Sadiq:2023zee} using previous versions of the method to investigate the LVK detected catalog, and also to demonstrate population estimation for ``light seed'' BH in mock LISA data~\cite{Sadiq:2024xsz}. The key development in the most recent version is to replace spherical (isotropic) kernels by elliptical or ellipsoidal, in other words allowing for independent choice or optimization of bandwidth along different dimensions of parameter space.  Rather than repeat the exposition of the complete method in~\cite{Sadiq:2025aog}, here, after explaining the general expression for the adaptive KDE of a given set of observations (events) $\vec{X}_i$, $i=1\ldots N$, we set out briefly how our application of the KDE addresses the fundamental statistical issues in population reconstruction for GW events, and summarize the expected strengths and weaknesses of the method. 

The estimated probability density $\hat{f}(\vec{x})$ at point $\vec{x}$ from a multidimensional Gaussian adaptive KDE is given by
\begin{multline}
 \label{eq:kdeatx}
    \hat{f}(\vec{x}) = \frac{1}{N \sqrt{(2\pi)^D|\Sigma|}} \sum_{i=1}^N \frac{1}{\lambda_i} \cdot \\
    \exp\left(-\frac{1}{2} (\vec{x} - \vec{X}_i)^T \lambda_i^{-2} \Sigma ^{-1} (\vec{x} - \vec{X_i})\right)
\end{multline}
where 
$D$ is the dimensionality of the data, $\Sigma$ is the global kernel covariance matrix (which we take to be the same for all data points), and the local adaptive parameter $\lambda_i$ is given by:
\begin{equation}
 \label{eq:lambdai}
 \lambda_i = \left( \frac{\hat{f}_0(\vec{X}_i)}{g} \right)^{-\alpha}, \, \, \log g = N^{-1} \sum_{i=1}^{N} \log \hat{f}_0(\vec{X}_i)\,.
\end{equation}
Here $\hat{f}_0$ is an initial pilot density estimate obtained by setting $\lambda_i=1$, $\alpha$ is the bandwidth sensitivity parameter (lying between 0 and 1), and $g$ is a normalization factor ensuring that the product of $\lambda_i$ over all events is unity. Our previous implementations were constrained by spherical kernels in standardized coordinates ($\Sigma = h^2\mathbb{I}$, where $\mathbb{I}$ is the identity matrix), however in \cite{Sadiq:2025aog} we added flexibility by allowing per-dimension bandwidths as $\Sigma = \mathrm{diag}(h_1^{-2}, h_2^{-2}, \ldots)$.  In principle a more general off-diagonal kernel matrix can be chosen, however our approach is intended to limit complexity and avoid the possibility that off-diagonal kernels might artificially enhance apparent correlations between parameters. 

\subsubsection{Bias-variance tradeoff: Bandwidth selection}
It is well known that bandwidth selection is crucial to control KDE errors and uncertainties \cite[e.g.][]{Silverman1986}, with too large bandwidth leading to a ``smooth'' but biased estimate which understates density variations, and conversely too small bandwidth leading to over-fitting with spuriously large fluctuations in the estimate and excessive variance over realizations.  Bandwidth choice is strongly linked to the finite statistics of detected events, since smaller bandwidth (in standardized coordinates) is appropriate for larger $N$.  We choose the multi-dimensional bandwidth $\vec{h} \equiv (h_1, h_2 \ldots)^T$ by maximizing the expected likelihood of the data set, evaluated by $k$-fold cross-validation (technically, the sum of log likelihoods over the $k$ folds).\footnote{Note that the distribution that maximizes the likelihood of the data $\vec{X}_i$ is a sum of delta functions \cite{Payne:2022xan}, comparable to the low-bandwidth limit of KDE which tends to infinite variance.}  The same optimization determines the adaptive parameter $\alpha$, which controls the scaling of the local bandwidth with the (pilot) density at the data points.  Adaptive KDE is crucial to obtaining a stable estimate with low bias given that the density of observations varies by orders of magnitude across the BBH parameter space~\cite{Sadiq:2021fin}. 

Given the high uncertainties in spin measurements, we choose to impose a minimum bandwidth over the (standardized) \cheff dimension of $0.2$, to mitigate cases where optimization appears to over-fit random sample fluctuations.  We find that this choice reduces statistical uncertainty relative to cases where smaller bandwidths are allowed, without adding noticeable bias.

\subsubsection{Uncertainty due to finite event statistics}
Given the relatively small number of GW observations, any estimate of the population distribution has a nontrivial uncertainty, characterized by considering the observed events as produced by an inhomogeneous Poisson process.  We estimate this uncertainty by constructing many bootstrap resampled data sets (during the iterative process described in Sec.~\ref{ss:iterkde}), which are intended to approximate the distribution of realizations of such a process~\cite{Sadiq:2023zee}.  At each iteration, we create a bootstrapped catalog by assigning a random weight of Poisson$(1)$ to each observed BBH; these weights are then applied when optimizing and evaluating the KDE, equivalent to taking Poisson$(1)$ copies of each event. We then quantify counting uncertainties via the median, symmetric $90\%$ intervals, etc., over several hundred bootstrap resampled KDEs. 

\subsubsection{Selection Function}
\label{subsec:selection}
The strategy described so far can naturally be applied to detected GW events, however these are a strongly biased sample of mergers in the local Universe.  More massive BBH events produce (up to some limit) significantly higher S/N than lower mass, and at constant mass higher $\cheff$ gives higher S/N \cite{Ajith:2009bn,Ng:2018neg}.  For our study, where any evolution over redshift is neglected (given the narrow range of $z$ currently accessible for the bulk of the BBH population~\cite{Sadiq:2025aog}), we quantify selection by the sensitive volume-time (VT) which relates the detected number density of events to the astrophysical merger rate density $\mathcal{R}$ as
\begin{equation}
\label{eq:VTdef}
  \D N_\mathrm{det}(m_1, m_2, \cheff) = \textnormal{VT}(m_1, m_2, \cheff) \D\mathcal{R}(m_1, m_2, \cheff).  
\end{equation}
We use an accurate fit to the probability of detection of simulated BBH signals in O3 data from~\cite{Lorenzo-Medina:2024opt}, using publicly released LVK analysis results~\cite{ligo_scientific_collaboration_and_virgo_2023_7890398,ligo_scientific_collaboration_and_virgo_2023_7890437};\footnote{These and other publicly released GWTC-3 data sets can be accessed at \url{https://gwosc.org/GWTC-3/}.}  we further integrate over luminosity distance, assuming a rate uniform in comoving volume-time, to compute VT at each point. 

We could also attempt to probe spin evolution over redshift by extending our KDE analysis to cover distance or redshift (as done for BBH masses in~\cite{Sadiq:2025aog}), rather than neglecting possible evolution.  However, our current methods require computational optimization for a complete 4-dimensional KDE analysis to be feasible.  As current data is inconclusive concerning evolution of the mass spectrum, we expect it to be similarly uninformative of spin evolution without imposing nontrivial model assumptions~(e.g.~\cite{Biscoveanu:2022qac}).

\subsubsection{Parameter measurement uncertainty}
\label{ss:iterkde}
In addition to the above effects, the properties of each event are uncertain due to detector noise, with most BBH having low S/N of order(10).  This uncertainty is quantified by providing ``PE'' samples, which represent the multi-dimensional Bayesian posterior density for each significant detected event~\cite{Veitch:2014wba,KAGRA:2021vkt,ligo_scientific_collaboration_and_virgo_2021_5546663}.  The standard posterior calculation uses an assumed uninformative prior probability over BH masses and spins: more specifically, a uniform distribution over redshifted apparent (``detector frame'') component masses and over spin magnitudes $|\vec{s}_{1,2}|$, and a isotropic distribution over spin directions.  As discussed in~\cite[e.g.][]{Mandel:2009pc,Thrane:2018qnx}, the resulting posteriors are only optimal and unbiased if the actual population distribution is equal to the prior.  In general, we will obtain more accurate parameter estimates by replacing the ``PE'' mass and spin prior by an estimate of the astrophysical population distribution. 

We implement this technically by \emph{iterative} density estimation with reweighting of the posterior samples~\cite{Sadiq:2023zee}.  
We here change our previous implementation of reweighting to reduce the effects of variance between bootstrap resamples.  Specifically, we now always use the mean over a buffer of 250 previous iterations to establish a population distribution for reweighting the next iteration: this ensures that such estimates are stable, and have support over the whole data set. 
To begin, we randomly select 250 posterior samples for each event
and build up a buffer of 250 KDEs, each using 1 randomly selected (unweighted) sample per event.  The resulting initial ``detected'' mean KDE neglects biases due to the PE prior and PE uncertainties and selection function, but is used as a well-behaved (low variance) starting point for subsequent reweighting.

\textrm{In subsequent iterations, we derive an estimated astrophysical distribution $f_\mathrm{pop}(m_1, m_2, \cheff)$ by dividing the current buffer mean KDE by $\textnormal{VT}(m_1, m_2, \cheff)$.  
We then randomly select one PE sample per event, not with uniform weights but reweighted by this population density divided by the PE prior over masses and over effective spin~\cite{Callister:2021gxf}.}\footnote{The mass prior is uniform except for the $(1+z)$ factor relating source to apparent ``detector frame'' masses.}  This reweighted sample set is used to generate a KDE, using Poisson bootstrap weights as described above; the new KDE is then appended to the buffer, while the ``oldest'' KDE in the buffer is removed. 

We thus create a slowly evolving buffer, which represents a (we hope) increasingly accurate estimate of the true detected population and its uncertainty.  If the buffer reaches an approximately stationary state, it will represent a self-consistent Monte Carlo estimate of the population and the parameters of each event~\cite{Sadiq:2023zee}. 
We monitor the approach to stationarity by examining ``time series'' over bootstrap iterations of the mean and standard deviation of the reweighted samples over each dimension, the optimized bandwidths, and the adaptive parameter, and calculating their autocorrelations.  We also monitor the effective number of independent PE samples for each detected event under the reweighting, to verify that it does not become very small (order $10$ or less) which would indicate inadequate sampling.  Typically these statistics appear stationary after 1000 bootstrap iterations with reweighting (i.e.\ 4 complete replacements of a buffer of length 250), and we then quote results using the subsequent 750 iterations.

\subsection{Tests and expected biases of iterative KDE reconstruction}
\label{ss:bias}
Simple 1d and 2d mock data tests of iterative KDE reconstruction, showing that it can properly account for PE uncertainty, have been presented in~\cite{Sadiq:2023zee}, and a detailed end-to-end mock data test on LISA ``light seed'' BBH sources was carried on in~\cite{Sadiq:2024xsz}, demonstrating its performance in the case of strongly correlated errors in the source total mass and redshift.  However, the iterative KDE has some inherent limitations in its ability to represent and reconstruct the true population, which should be understood before interpreting our results. 

\subsubsection{Parameter resolution and scale of density variations} 
The most obvious restriction is that the Gaussian KDE cannot represent abrupt, step-function-like variations in density over the source parameter space (see~\cite{Sadiq:2023zee} for a mock data example).  The impact of this restriction to a ``smooth'' functional form also varies with the local density of detected events, due to the adaptive kernel choice Eq.~\eqref{eq:lambdai}: the (log of the) estimated density may vary rapidly in regions with a high number of detections, but must be slowly varying in sparsely populated regions.  While this is desirable to prevent over-fitting and excess variance, it limits the possible parameter dependence of the estimated population.  In particular, if the true population does contain abrupt steps or very narrow peak features, we expect the KDE to provide an ``over-smoothed'' reconstruction, and also that the amount of excess smoothing will decrease (i.e., the resolution of the estimate will improve) with higher numbers of detected events. 

\subsubsection{Behavior in regions with low event count} 
A related but distinct limitation concerns the limit of the KDE in regions far from most detected events, where a small number of kernels (possibly only a single kernel) contribute non-negligibly.  Here the estimate for the distribution of detected events must go towards zero with a Gaussian dependence, and will be dominated at large distance by the kernel with highest bandwidth.  The consequences for the astrophysical population estimate will depend on the parameter dependence of the selection function, as discussed in more detail in~\cite{Sadiq:2024xsz,Sadiq:2025aog} for the cases of binary masses and redshift.  Considering the space of binary masses and spins, the selection function $\textnormal{VT}$ typically varies as a power law or polynomial~\cite{Fishbach:2018edt,Lorenzo-Medina:2024opt}, thus the estimated astrophysical distribution far from the detected signal points will also be dominated by the Gaussian falloff of the ``broadest kernel''.  The uncertainties in rate density for points where only order(1) detected events contribute will also be extremely large (typically orders of magnitude), as reflected in the bootstrap estimates.

\section{Application to GWTC-3 Data}

\begin{figure}
    \centering
    \includegraphics[width=0.9\columnwidth]{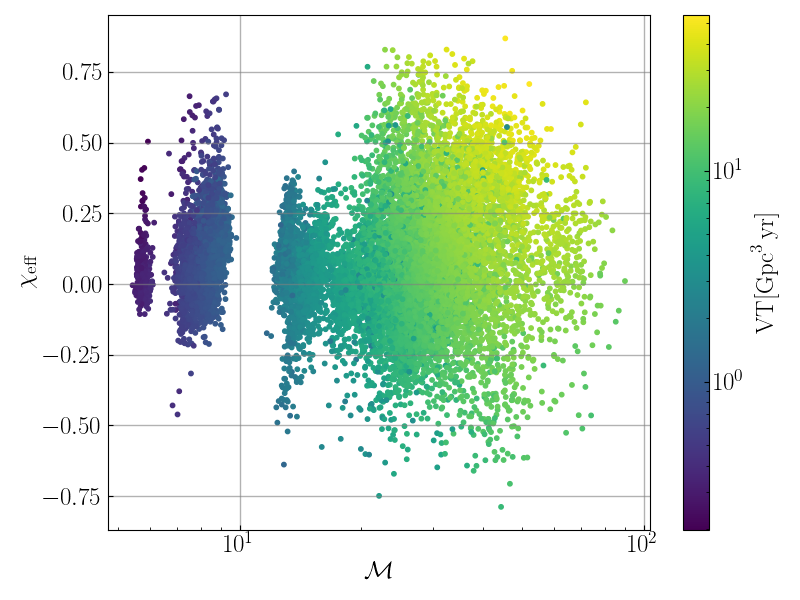}
    \caption{Chirp mass $\mathcal{M}$ and \cheff of PE samples for 69 BBH events from GWTC-3, 
    with color showing the sensitive volume-time (VT). 
}
    \label{fig:pe-samples-VT}
\end{figure}

We apply the iterative reweighted KDE to GWTC-3 public PE samples \cite{ligo_scientific_collaboration_and_virgo_2021_5546663,ligo_scientific_collaboration_and_virgo_2021_5117970}, selecting 69 BBH candidate events from the cumulative O1-O3 catalog which have a false alarm rate below 1 per year and excluding GW190814, an outlier event with an unusually low secondary mass ($m_2\simeq 2.6\,\msun$)~\cite{LIGOScientific:2020zkf}, 
as this system could represent either a very massive neutron star or a light black hole, which is inconsistent with the primary BBH population we aim to analyze.  
For the highest-mass BBH GW190521, we prefer available PE results using the \textsc{NRSurd7q4} waveform model \cite{190521_pe} which is expected to give more accurate estimates than standard GWTC-3 choices for this event~\cite{LIGOScientific:2020iuh}.
Drawing 250 random samples from the posterior for each BBH event, 
our sensitive $\textnormal{VT}$ estimates at the masses and \cheff of the samples~\cite{Lorenzo-Medina:2024opt} are shown in Fig.~\ref{fig:pe-samples-VT}. 

We perform KDEs over the logarithms of source component masses $\ln m_1$, $\ln m_2$, and $\cheff$, ensuring proper exchange symmetry by ``reflecting'' each $(\ln m_1, \ln m_2)$ sample point, i.e.\ also including the mass-exchanged point $(\ln m'_1=\ln m_2, \ln m'_2=\ln m_1)$ in the KDE data~\cite{Sadiq:2023zee}.  Having obtained 750 bootstrap estimates of the detected distribution, we evaluate these estimates on a 3d grid for visualization, and convert to astrophysical merger rate density by multiplying by the total number of detected events and dividing by VT evaluated on the same 3d grid. 

\subsection{Marginalized mass distributions}

From the 3d rate estimates, we first marginalize over the $\cheff$ dimension to obtain the merger rate as a function of binary component masses. 
\begin{figure}[tbp]
    \centering
    \includegraphics[width=0.99\columnwidth]{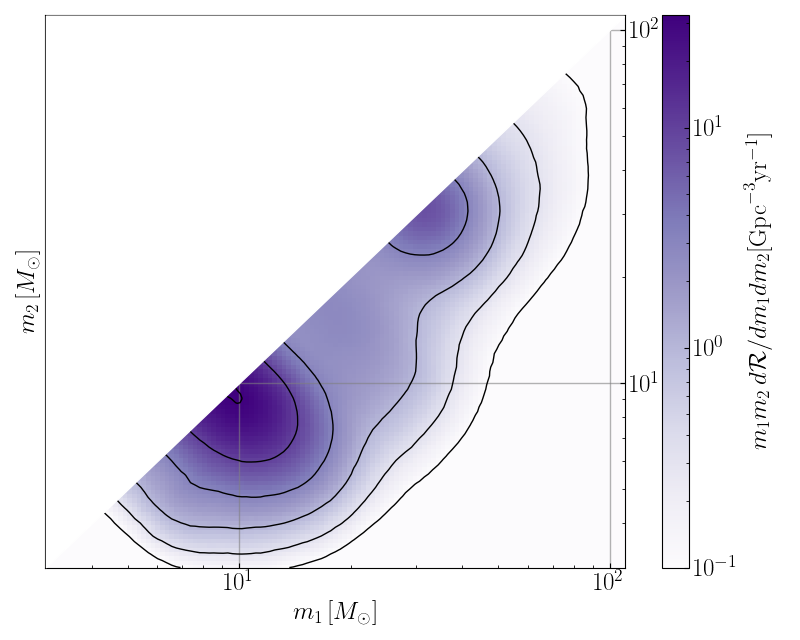}
    \caption{Merger rate density over binary component masses, marginalized over $\cheff$. Median estimate over bootstrap iterations.}
    \label{fig:2Dmass-rate}
\end{figure}
The resulting two-dimensional mass distribution in Fig.~\ref{fig:2Dmass-rate} is similar to that found under the assumptions previously made in \cite{Sadiq:2023zee}, which neglected spin in selection effects and effectively assumed the population \cheff distribution to be equal to the PE prior (i.e.\ symmetric around zero).  In particular, the primary and secondary masses appear anti-correlated in the neighborhood of the $\sim\!10\,\msun$ peak, with significant support for unequal mass binaries: contours of constant rate density are approximate ellipses with major axis parallel to a line $m_1m_2 = \textnormal{const.}$.  Conversely, the $\sim\!30\,\msun$ peak is concentrated near the equal mass line (constant rate contours are near-circles centered at $m_1=m_2$), and the secondary mass distribution declines rapidly above $\sim\!35\,\msun$.  Minor differences from our previous results appear in the high-mass regime.
Additionally, overall rates are somewhat lower than under our previous zero-spin assumption, possibly reflecting the influence of aligned spin on detection probability and thus on VT. 

\begin{figure}[tbp]
    \centering
    \hspace*{-0.2cm}
    \includegraphics[width=\columnwidth]{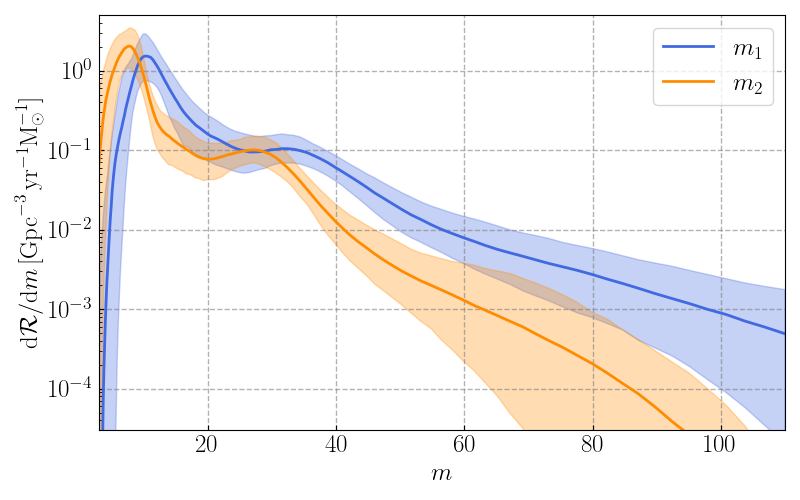}
    \caption{Rate density over BBH component masses obtained by marginalizing 3d rate estimates over \cheff and one component mass. Color bands show 90\% confidence regions obtained from bootstrap iterations.}
    \label{fig:Rate-oneD-masses}
\end{figure}

We further marginalize over each component mass component to obtain one-dimensional merger rates over primary and secondary masses, shown in Fig.~\ref{fig:Rate-oneD-masses}. 

The resulting 1d rates are consistent with our previous analysis, 
but again with minor overall shifts 
which likely reflect the inclusion of aligned spin dependence in VT. 

\subsection{Dependence of aligned spin on binary masses}

\begin{figure}[tbp]
    \centering
    \hspace*{-0.2cm}
    \includegraphics[width=1.03\columnwidth]{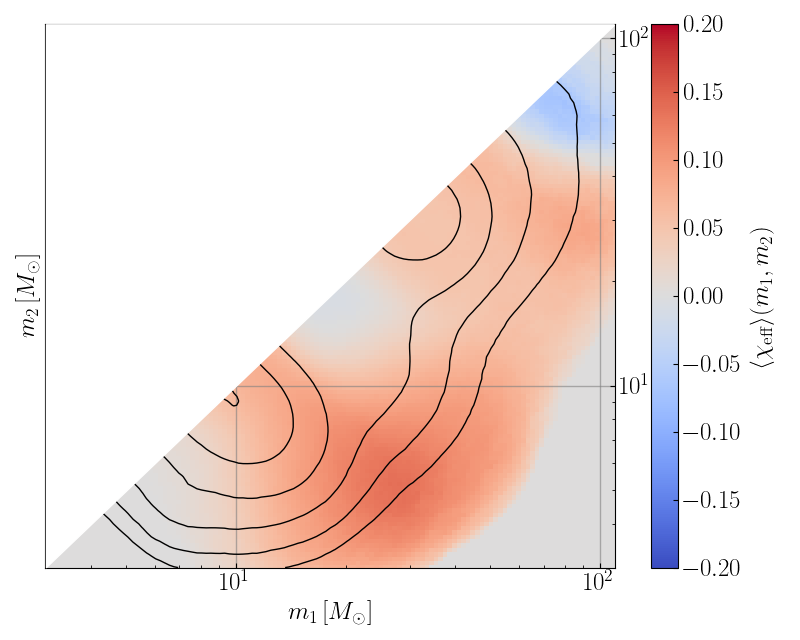}
    \caption{Population mean \cheff as a function of BH component masses (median estimate over bootstrap iterations).  Contours show rate density as in Fig.~\ref{fig:2Dmass-rate}.}
    \label{fig:2Dmass-meancf}
\end{figure}

Although it is impossible in practice to visualize the full 3d mass-spin distribution with uncertainties, we start by characterizing the dependence of basic statistical properties of the population \cheff distribution as functions of $(m_1, m_2)$.  We show the \emph{population mean} \cheff (median estimate over bootstrap iterations) over the 2d mass plane in Fig.~\ref{fig:2Dmass-meancf}. 
We see a number of interesting features: the local maximum of merger rate near $m_1=m_2=30\,\msun$ has small but marginally positive $\langle\cheff\rangle$, while the region around $m_1 \simeq m_2 \simeq 20\,\msun$ has zero population mean.  Conversely, the distribution at lower secondary mass shows an apparent trend towards higher $\langle\cheff\rangle$ at unequal mass, reflecting the correlation found in \cite{Callister:2021fpo}. 
Note that this trend of increasing \cheff with increasing (decreasing) $m_1$ ($m_2$) for low-mass BBH was present in the original unprocessed PE samples, where it is due to degeneracies in BBH waveforms~\cite{Baird:2012cu}.  Thus, we cannot rule out that the apparent population trend is partly due to correlated PE uncertainty, which the reweighting method may not be able to completely eliminate. 
At high masses and far from the detected events the population trends towards negative or near-zero $\langle\cheff\rangle$, which may primarily reflect random variation in regions with small number statistics. 

\begin{figure}[tbp]
    \centering
    \includegraphics[width=1.0\columnwidth]{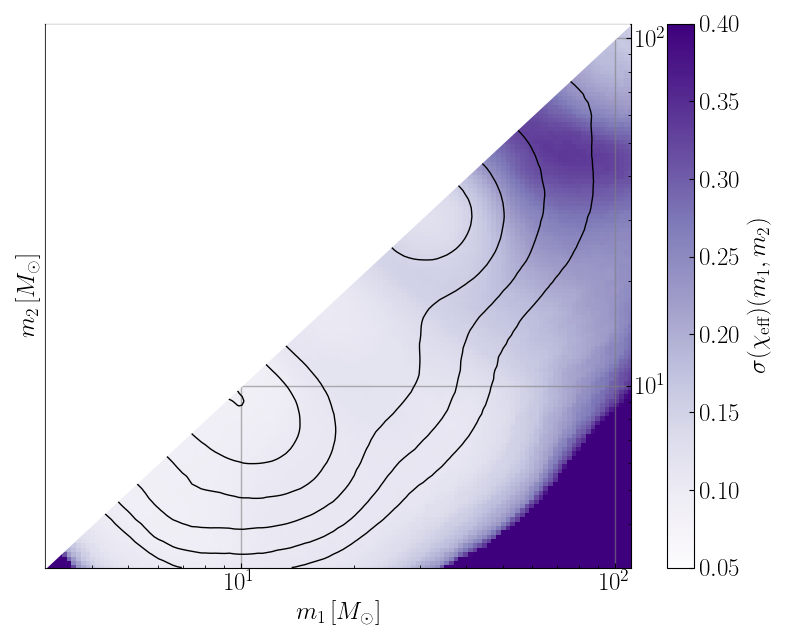}
    \caption{Population standard deviation of \cheff as a function of BH component masses (median estimate over bootstrap iterations).  Contours show rate density as in Fig.~\ref{fig:2Dmass-rate}.}
    \label{fig:2Dmass-stdcf}
\end{figure}

We further show the \emph{population standard deviation} of \cheff in Fig.~\ref{fig:2Dmass-stdcf}, as a measure of the spread of the spin distribution. 
This measure remains small over most of the parameter space except for primary mass above $\sim\!40\,\msun$ and secondary above $\sim\!25\,\msun$, where we observe a rapid increase in $\sigma(\cheff)$.  (The standard deviation also becomes large far from the detected events, where small number statistics dominate.)  An increase in the width of the \cheff distribution at higher masses was proposed in \cite{Baibhav:2020xdf,Antonini:2024het} as a potential signature of hierarchical BBH formation, if 1st generation BH 
are assumed to have masses limited by PISN dynamics
and to have small spin magnitudes, 
and given that 2nd generation BH have spin magnitudes of order $0.7$.  We observe a relatively sharp transition to a broad spin distribution above $m_1\simeq 40\,\msun$; the coincidence of this approximate mass scale with stellar evolution estimates of the PISN ``gap'' was noted in~\cite{Antonini:2024het} as further motivating dynamic and hierarchical formation for heavier BH in merging binaries. 

\subsection{Mass and effective spin}

\begin{figure}[tbp]
    \centering
    \includegraphics[width=0.85\columnwidth]{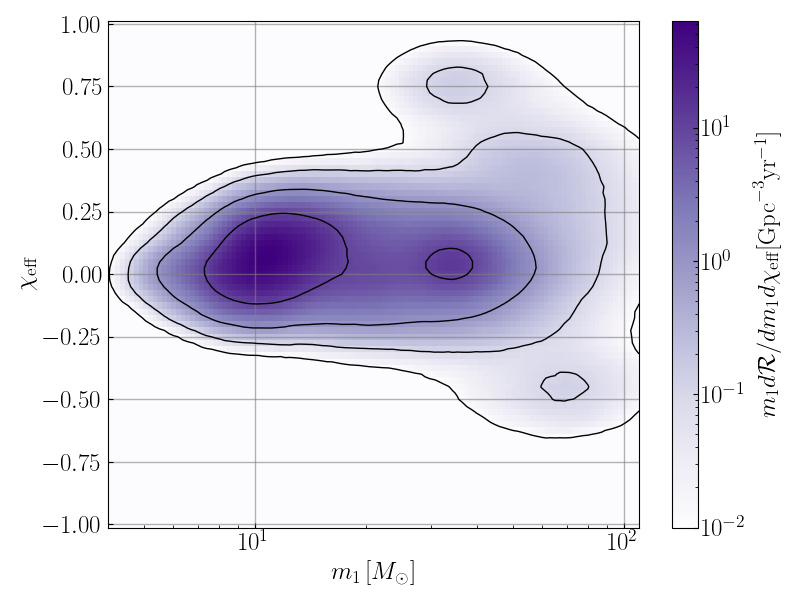}
    \\
    \includegraphics[width=0.85\columnwidth]{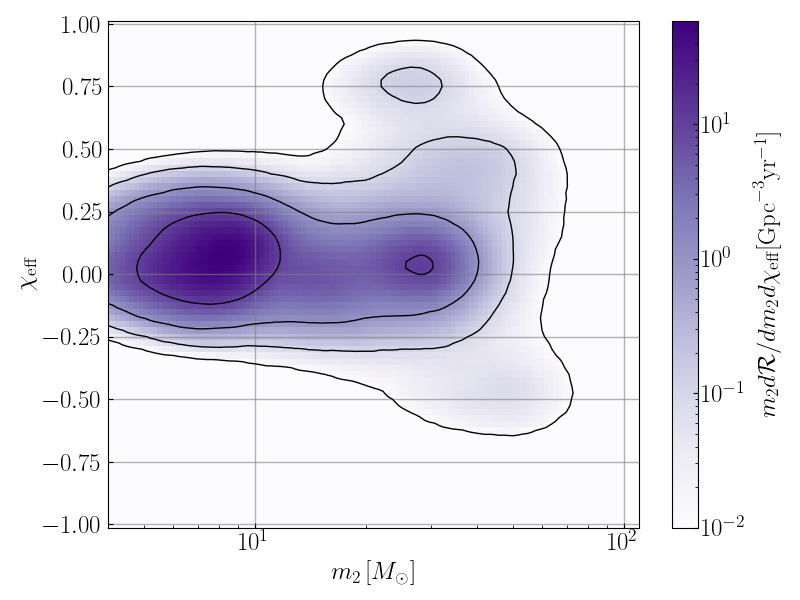}
    \caption{2d rate density over binary component mass and \cheff (median estimate over bootstrap iterations).  Top (bottom): marginalized over $m_2$ ($m_1$).}
    \label{fig:mass-Xieff-rate}
\end{figure}

\begin{figure}[tbp]
    \centering
    \includegraphics[width=0.75\columnwidth]{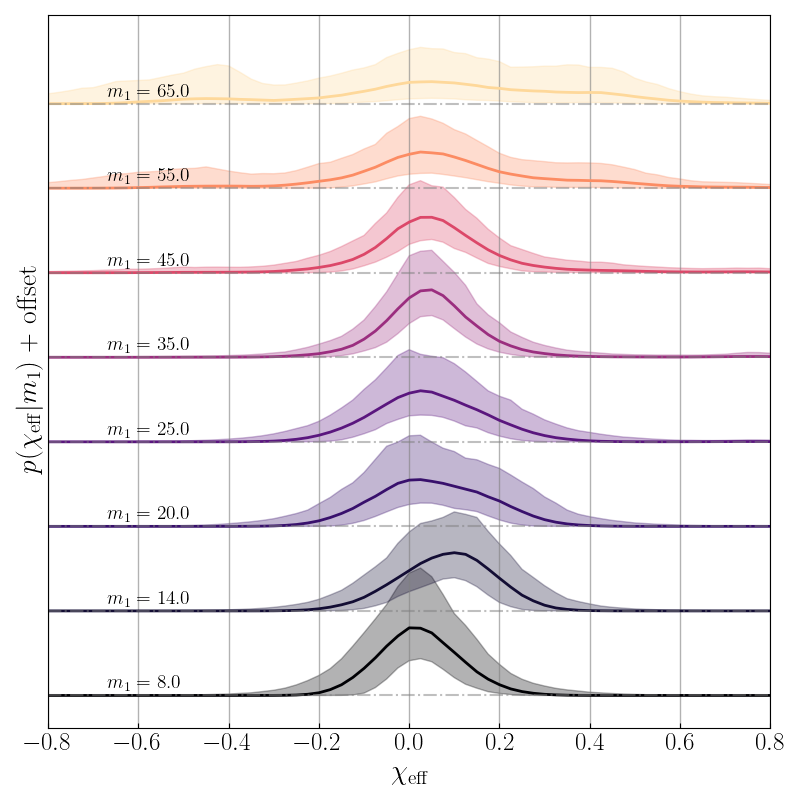} \\
    \includegraphics[width=0.75\columnwidth]{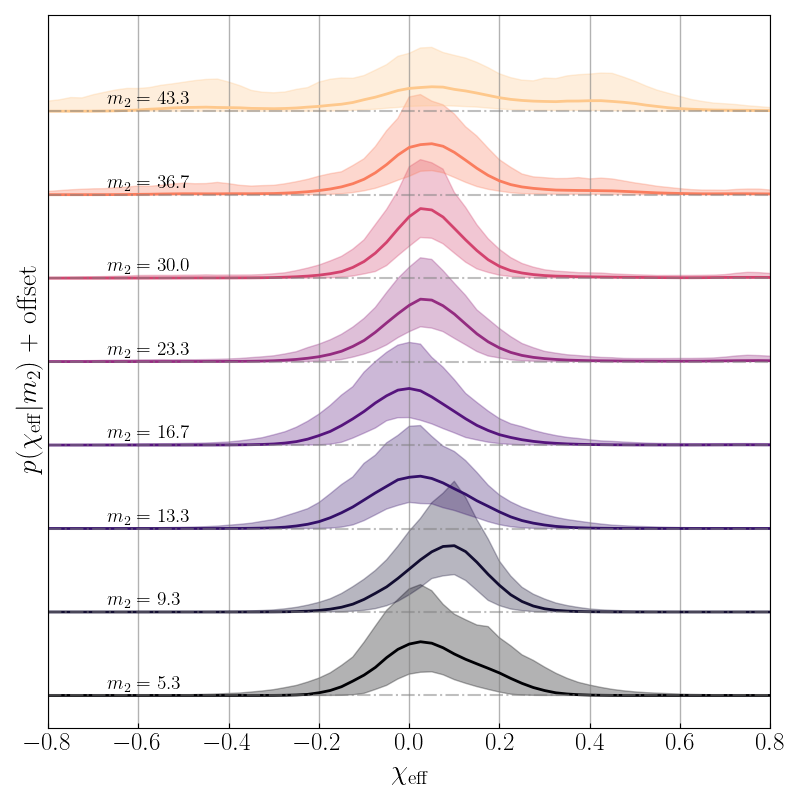}
    \caption{Normalized merger rate distributions over effective spin at constant component mass.  Top: Population distribution of \cheff marginalized over $m_2$ at constant $m_1$ values.  Bottom: Distribution of \cheff marginalized over $m_1$ at constant $m_2$.  We show the median and symmetric 90\% confidence intervals over bootstrap iterations and apply constant vertical offsets for visual clarity.}
    \label{fig:chieffoffset}
\end{figure}

We may isolate trends over primary or secondary mass by plotting the \cheff distribution marginalizing over the other component mass, as shown in Figure~\ref{fig:mass-Xieff-rate}. 
%
We see a strong concentration of events at low to intermediate masses ($\sim\!10$ -- $35\,\msun$) and near-zero effective spin (slightly positive for the $\sim\!10\,\msun$ peak), consistent with previous non-parametric analyses \cite{Ray:2024hos,Heinzel:2024hva}.  However, we also observe nontrivial additional structure at higher masses, 
with a larger general spread of non-zero spin values for $m_1\gtrsim 40\,\msun$, $m_2>30\,\msun$, and hints (at low significance) of a small sub-population with high positive \cheff around $m_1\simeq 35\,\msun$.  The peaks near $\cheff = 0$ appear to co-exist with this broader sub-population over some range of mass.  The significance of these trends can be visualized by selecting discrete component mass values and computing the normalized distribution of \cheff for each, as in Fig.~\ref{fig:chieffoffset}.
We indeed see an increase in the spread of the \cheff distribution as $m_1$ increases from $35\,\msun$ through $55\,\msun$, although statistical uncertainties dominate at $65\,\msun$ (and higher masses which we do not plot).  Significant trends over $m_2$ are, though, harder to identify; this difference may be expected as the secondary BH makes a smaller contribution to \cheff simply by construction. 

The low-mass region also shows interesting structure: the mean of \cheff shifts from near zero at the lowest masses to clearly positive at $m_1\sim 14\,\msun$, then closer to zero at $m_1\sim 20\,\msun$ and above, a trend also visible in Fig.~\ref{fig:2Dmass-meancf}.  It is believed that the low-mass peak is associated with isolated binary formation~\cite{Antonini:2022vib,Godfrey:2023oxb}, for which mass transfer effects (either stable or common-envelope) may play important roles in determining BH spins~\cite[e.g.][]{Mould:2022xeu,Olejak:2024qxr}; stable mass transfer~\cite{2022ApJ...940..184V} has been discussed as a possible origin for correlations between binary mass ratio and aligned spin~\cite{Broekgaarden:2022nst,Banerjee:2024wbq}. 
Alternatively, more complex scenarios where different sub-populations coexist even at low mass may be responsible for such trends~\cite{Galaudage:2024meo}.  While our findings do not match any specific model, the ability of our 3d analysis to localize mass-spin correlations in parameter space may provide a crucial guide to identify the relevant effects. 

\begin{figure}
\hspace*{-0.48cm}
\includegraphics[width=0.515\columnwidth]{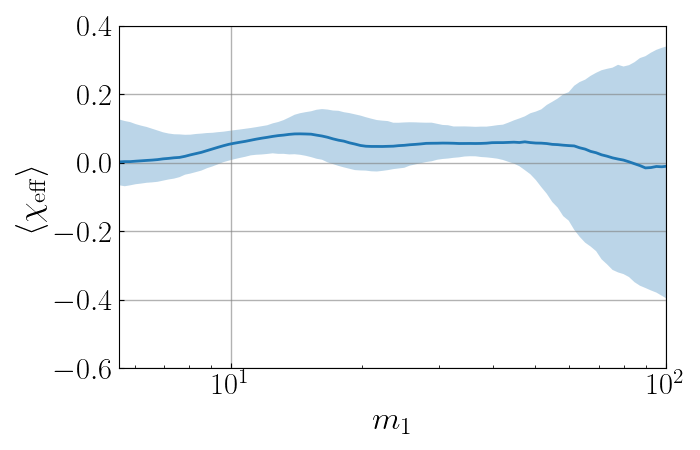}
\hspace*{-0.34cm}
\includegraphics[width=0.515\columnwidth]{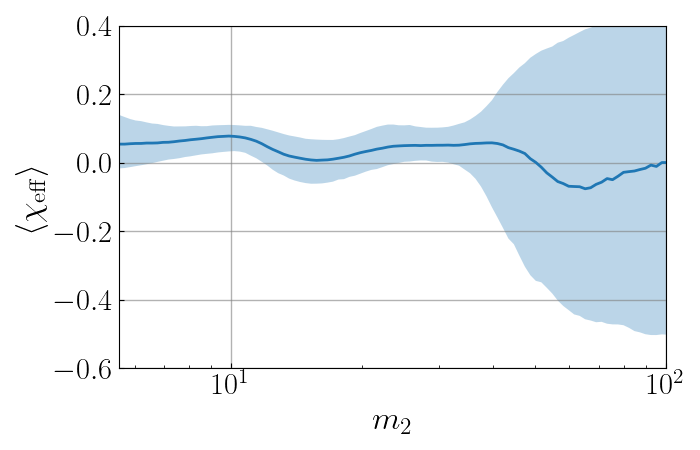} \\
\hspace*{-0.30cm}
\includegraphics[width=0.5\columnwidth]{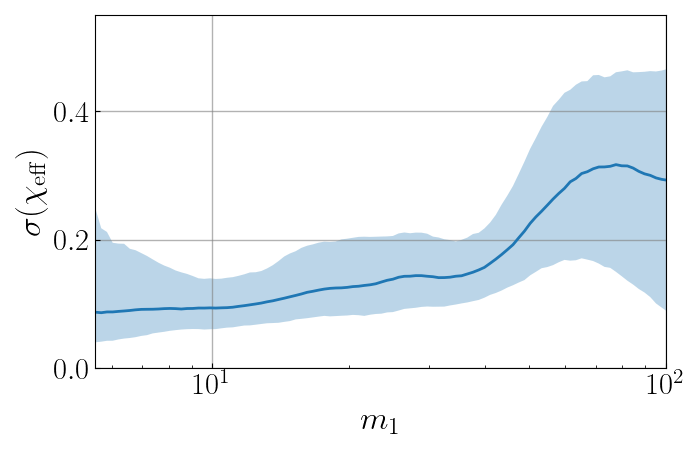}
\hspace*{-0.22cm}
\includegraphics[width=0.5\columnwidth]{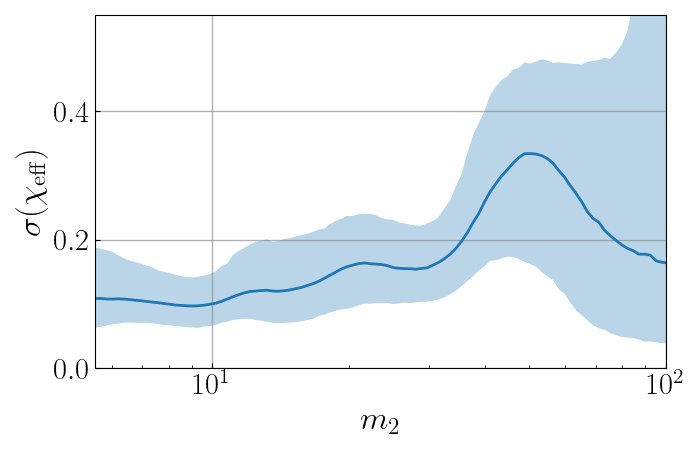}
    \caption{The mean (top panels) and standard deviation (bottom panels) of 
    the population \cheff distribution, as a function of primary (left) or secondary (right) mass (median and symmetric 90\% confidence intervals over bootstrap iterations).}
    \label{fig:chieff-mean-std}
\end{figure}

Lastly, we further investigate the significance of variations in \cheff by plotting the dependence of its population statistics (mean and standard deviation) against component masses, as in Fig.~\ref{fig:chieff-mean-std}. 
The population mean indeed takes significantly positive (though small) values for $m_1 \gtrsim 10\,\msun$ and low $m_2$, but is generally consistent with $0$ for higher mass, where statistical uncertainty grows rapidly.  On the other hand the dispersion (spread) remains small at low mass and increases at $m_1 > 45\,\msun$ ($m_2 > 35\,\msun$), confirming findings from the parameterized analysis of~\cite{Antonini:2024het}. 

We caution that separating real population features from statistical artifacts requires understanding of potential biases in the reconstruction method, as mentioned in Sec.~\ref{ss:bias}.  In particular, apparent broadening of the \cheff distribution may be partly caused by increased kernel bandwidth in regions with few detections, or by PE uncertainties which tend to grow steadily with increasing mass. However, we do not expect such effects to give rise to the specific mass dependence apparent in our results. 
Any statistical biases will also be reduced using future event catalogs, with more precise spin measurements (due to higher SNRs) and higher detection counts.

\section{Discussion}

We argue that maximally flexible or data-driven methods are preferable for investigating binary black hole population properties based on GW observations, as there is currently no clear preferred astrophysical explanation of BBH formation, and restrictive assumptions may induce biases in inference.  In this work we take a further step towards a general non-parametric analysis with a reconstruction of the population from GWTC-3 detections over the 3d space of component masses and effective aligned spin, arguably the parameters holding the most information on possible correlations or subpopulations (for the current set of detections confined to relatively low $z$). 
We employ an iterative KDE-based method designed to account for finite event statistics, parameter uncertainties and selection effects~\cite{Sadiq:2023zee,Sadiq:2024xsz,Sadiq:2025aog}.


For the BBH mass distribution, we obtain similar findings to our previous work neglecting spin effects~\cite{Sadiq:2023zee}, indicating features (comparable to~\cite{Tiwari:2021yvr}) that depart from the common assumption of a single population distribution over mass ratio.  

Our reconstruction of the population over aligned spin shows bulk features similar to the LVK analyses~\cite{LIGOScientific:2020kqk,KAGRA:2021duu}: the majority of the population lies close to zero \cheff, with a preference for small positive values in regions of the mass space where the distribution is narrowly defined.  Both the dispersion of the \cheff distribution and its measurement uncertainties grow noticeably for component masses above $\sim\!40\,\msun$ (comparable to \cite{Tiwari:2021yvr,Biscoveanu:2022qac}).  Despite such high uncertainties, we find a significant increase in the \cheff dispersion for primary mass between $40-50\,\msun$, comparable to the transition to a broader distribution consistent with a zero mean at higher masses which is claimed in~\cite{Antonini:2024het} to be due to hierarchical merger dominating above the expected PISN mass scale of $\sim\!45\,\msun$. 

We partly confirm the mass ratio-\cheff correlation claimed in~\cite{Callister:2021fpo,Adamcewicz:2023mov}, as we find a positive mean \cheff above $\sim\!0.1$ only outside the near-equal-mass region inhabited by the bulk of detected events.  However, this positive aligned-spin trend does not persist to high-mass asymmetric binaries; this restriction is consistent with the possibility that such correlations are due to isolated binary dynamics~\cite[e.g.][]{Banerjee:2024wbq}. 

We identify other trends such as an apparent positive correlation between \cheff and $m_1$ in the low-mass BBH rate peak, possibly linked to mass transfer: overall, while these features offer hints to formation channels and guidance for astrophysical modeling, their statistical significance is not high, underscoring the need for future data to refine such interpretations.  We thus expect current and upcoming LVK O4 catalog releases to have a decisive role in confirming or ruling out claimed correlations and subpopulations. 

Our KDE-based analysis, while flexible and computationally efficient, has some limitations. The assumption of a smooth continuous kernel reduces our ability to identify ``sharp'' structures in the underlying distribution, if such exist, while the sheer size of the 3d parameter space requires a large number of detections to obtain useful ``local'' measurements.  Additionally, this study focused solely on effective spins (\cheff), leaving in-plane spin components and spin magnitudes (not to mention possible evolution over redshift~\cite{Biscoveanu:2022qac}) as critical future extensions.  As the GW catalog grows, combining our method with improved astrophysical modeling will be crucial to disentangling the complex origins of merging black holes (e.g.~\cite{Fabbri:2025faf} discusses comparison of non-parametric inference with physical models). 

\acknowledgments
The authors have benefited from conversations with Fabio Antonini, Vaibhav Tiwari, Steve Fairhurst, Will Farr and others in the LVK Binary Rates \& Populations group. 
JS also acknowledges support from the European Union’s H2020 ERC Consolidator Grant ``GRavity from Astrophysical to Microscopic Scales'' (GRAMS-815673) and the EU Horizon 2020 Research and Innovation Programme under the Marie Sklodowska-Curie Grant Agreement No.\ 101007855. 
TD and ALM acknowledge support from the Grant ED431F 2025/04 of the Galician Conselleria de Educacion, Ciencia, Universidades e Formacion Profesional, have received financial support from Xunta de Galicia (CIGUS Network of research centers) and the European Union, and are supported by Mar{\' i}a de Maeztu grant CEX2023-001318-M funded by MICIU/AEI/10.13039/501100011033.

The authors are grateful for computational resources provided by the LIGO Laboratory and supported by National Science Foundation Grants PHY-0757058 and PHY-0823459. This research has made use of data or software obtained from the Gravitational Wave Open Science Center (gwosc.org), a service of the LIGO Scientific Collaboration, the Virgo Collaboration, and KAGRA. This material is based upon work supported by NSF's LIGO Laboratory which is a major facility fully funded by the National Science Foundation, as well as the Science and Technology Facilities Council (STFC) of the United Kingdom, the Max-Planck-Society (MPS), and the State of Niedersachsen/Germany for support of the construction of Advanced LIGO and construction and operation of the GEO600 detector. Additional support for Advanced LIGO was provided by the Australian Research Council. Virgo is funded, through the European Gravitational Observatory (EGO), by the French Centre National de Recherche Scientifique (CNRS), the Italian Istituto Nazionale di Fisica Nucleare (INFN) and the Dutch Nikhef, with contributions by institutions from Belgium, Germany, Greece, Hungary, Ireland, Japan, Monaco, Poland, Portugal, Spain. KAGRA is supported by Ministry of Education, Culture, Sports, Science and Technology (MEXT), Japan Society for the Promotion of Science (JSPS) in Japan; National Research Foundation (NRF) and Ministry of Science and ICT (MSIT) in Korea; Academia Sinica (AS) and National Science and Technology Council (NSTC) in Taiwan.

\bibliography{reference}

@article{Sadiq:2021fin,
    author = "Sadiq, Jam and Dent, Thomas and Wysocki, Daniel",
    title = "{Flexible and fast estimation of binary merger population distributions with an adaptive kernel density estimator}",
    eprint = "2112.12659",
    archivePrefix = "arXiv",
    primaryClass = "gr-qc",
    doi = "10.1103/PhysRevD.105.123014",
    journal = "Phys. Rev. D",
    volume = "105",
    number = "12",
    pages = "123014",
    year = "2022"
}

@article{Sadiq:2023zee,
    author = "Sadiq, Jam and Dent, Thomas and Gieles, Mark",
    title = "{Binary Vision: The Mass Distribution of Merging Binary Black Holes via Iterative Density Estimation}",
    eprint = "2307.12092",
    archivePrefix = "arXiv",
    primaryClass = "astro-ph.HE",
    doi = "10.3847/1538-4357/ad0ce6",
    journal = "Astrophys. J.",
    volume = "960",
    number = "1",
    pages = "65",
    year = "2024"
}

@article{Sadiq:2024xsz,
    author = "Sadiq, Jam and Dey, Kallol and Dent, Thomas and Barausse, Enrico",
    title = "{Reconstructing the LISA massive black hole binary population via iterative kernel density estimation}",
    eprint = "2410.17056",
    archivePrefix = "arXiv",
    primaryClass = "gr-qc",
    doi = "10.1103/PhysRevD.111.063051",
    journal = "Phys. Rev. D",
    volume = "111",
    number = "6",
    pages = "063051",
    year = "2025"
}

@article{Sadiq:2025aog,
      title={Looking To The Horizon: Probing Evolution in the Black Hole Spectrum With GW Catalogs}, 
      author={Jam Sadiq and Thomas Dent and Ana Lorenzo Medina},
      journal={},
      year={2025},
      eprint={2502.06451},
      archivePrefix={arXiv},
      primaryClass={gr-qc},
      url={https://arxiv.org/abs/2502.06451}, 
}

@article{Lorenzo-Medina:2024opt,
    author = "Lorenzo-Medina, Ana and Dent, Thomas",
    title = "{A physically modelled selection function for compact binary mergers in the LIGO-Virgo O3 run and beyond}",
    eprint = "2408.13383",
    archivePrefix = "arXiv",
    primaryClass = "gr-qc",
    doi = "10.1088/1361-6382/ad9c0e",
    journal = "Class. Quant. Grav.",
    volume = "42",
    number = "4",
    pages = "045008",
    year = "2025"
}

@BOOK{Silverman1986,
	title = {Density Estimation for Statistics and Data Analysis},
	publisher = {Chapman and Hall, London},
	author = {B.W. Silverman},
	year = {1986},
	edition = {1st},
}

@article{LIGOScientific:2016vpg,
    author = "Abbott, B. P. and others",
    collaboration = "LIGO Scientific, Virgo",
    title = "{Astrophysical Implications of the Binary Black-Hole Merger GW150914}",
    eprint = "1602.03846",
    archivePrefix = "arXiv",
    primaryClass = "astro-ph.HE",
    reportNumber = "LIGO-P1500262",
    doi = "10.3847/2041-8205/818/2/L22",
    journal = "Astrophys. J. Lett.",
    volume = "818",
    number = "2",
    pages = "L22",
    year = "2016"
}

@article{LIGOScientific:2020kqk,
    author = "Abbott, R. and others",
    collaboration = "LIGO Scientific, Virgo",
    title = "{Population Properties of Compact Objects from the Second LIGO-Virgo Gravitational-Wave Transient Catalog}",
    eprint = "2010.14533",
    archivePrefix = "arXiv",
    primaryClass = "astro-ph.HE",
    reportNumber = "LIGO-P2000077",
    doi = "10.3847/2041-8213/abe949",
    journal = "Astrophys. J. Lett.",
    volume = "913",
    number = "1",
    pages = "L7",
    year = "2021"
}

@article{LIGOScientific:2016aoc,
    author = "Abbott, B. P. and others",
    collaboration = "LIGO Scientific, Virgo",
    title = "{Observation of Gravitational Waves from a Binary Black Hole Merger}",
    eprint = "1602.03837",
    archivePrefix = "arXiv",
    primaryClass = "gr-qc",
    reportNumber = "LIGO-P150914",
    doi = "10.1103/PhysRevLett.116.061102",
    journal = "Phys. Rev. Lett.",
    volume = "116",
    number = "6",
    pages = "061102",
    year = "2016"
}

@article{LIGOScientific:2018mvr,
    author = "Abbott, B. P. and others",
    collaboration = "LIGO Scientific, Virgo",
    title = "{GWTC-1: A Gravitational-Wave Transient Catalog of Compact Binary Mergers Observed by LIGO and Virgo during the First and Second Observing Runs}",
    eprint = "1811.12907",
    archivePrefix = "arXiv",
    primaryClass = "astro-ph.HE",
    reportNumber = "LIGO-P1800307",
    doi = "10.1103/PhysRevX.9.031040",
    journal = "Phys. Rev. X",
    volume = "9",
    number = "3",
    pages = "031040",
    year = "{2019}"
}

@article{LIGOScientific:2020ibl,
    author = "Abbott, R. and others",
    collaboration = "LIGO Scientific, Virgo",
    title = "{GWTC-2: Compact Binary Coalescences Observed by LIGO and Virgo During the First Half of the Third Observing Run}",
    eprint = "2010.14527",
    archivePrefix = "arXiv",
    primaryClass = "gr-qc",
    reportNumber = "P2000061",
    doi = "10.1103/PhysRevX.11.021053",
    journal = "Phys. Rev. X",
    volume = "11",
    pages = "021053",
    year = "{2021}"
}

@article{KAGRA:2021vkt,
    author = "Abbott, R. and others",
    collaboration = "KAGRA, VIRGO, LIGO Scientific",
    title = "{GWTC-3: Compact Binary Coalescences Observed by LIGO and Virgo during the Second Part of the Third Observing Run}",
    eprint = "2111.03606",
    archivePrefix = "arXiv",
    primaryClass = "gr-qc",
    reportNumber = "LIGO-P2000318",
    doi = "10.1103/PhysRevX.13.041039",
    journal = "Phys. Rev. X",
    volume = "13",
    number = "4",
    pages = "041039",
    year = "2023"
}

@article{KAGRA:2021duu,
    author = "Abbott, R. and others",
    collaboration = "KAGRA, VIRGO, LIGO Scientific",
    title = "{Population of Merging Compact Binaries Inferred Using Gravitational Waves through GWTC-3}",
    eprint = "2111.03634",
    archivePrefix = "arXiv",
    primaryClass = "astro-ph.HE",
    reportNumber = "LIGO-P2100239 ; Data release: https://zenodo.org/record/5655785, LIGO-P2100239",
    doi = "10.1103/PhysRevX.13.011048",
    journal = "Phys. Rev. X",
    volume = "13",
    number = "1",
    pages = "011048",
    year = "2023"
}

@article{Ajith:2009bn,
    author = "Ajith, P. and others",
    title = "{Inspiral-merger-ringdown waveforms for black-hole binaries with non-precessing spins}",
    eprint = "0909.2867",
    archivePrefix = "arXiv",
    primaryClass = "gr-qc",
    doi = "10.1103/PhysRevLett.106.241101",
    journal = "Phys. Rev. Lett.",
    volume = "106",
    pages = "241101",
    year = "2011"
}

@article{Zevin:2022wrw,
    author = "Zevin, Michael and Bavera, Simone S.",
    title = "{Suspicious Siblings: The Distribution of Mass and Spin across Component Black Holes in Isolated Binary Evolution}",
    eprint = "2203.02515",
    archivePrefix = "arXiv",
    primaryClass = "astro-ph.HE",
    doi = "10.3847/1538-4357/ac6f5d",
    journal = "Astrophys. J.",
    volume = "933",
    number = "1",
    pages = "86",
    year = "2022"
}

@article{Broekgaarden:2022nst,
    author = "Broekgaarden, Floor S. and Stevenson, Simon and Thrane, Eric",
    title = "{Signatures of Mass Ratio Reversal in Gravitational Waves from Merging Binary Black Holes}",
    eprint = "2205.01693",
    archivePrefix = "arXiv",
    primaryClass = "astro-ph.HE",
    doi = "10.3847/1538-4357/ac8879",
    journal = "Astrophys. J.",
    volume = "938",
    number = "1",
    pages = "45",
    year = "2022"
}

@article{Fuller:2019sxi,
    author = "Fuller, Jim and Ma, Linhao",
    title = "{Most Black Holes are Born Very Slowly Rotating}",
    eprint = "1907.03714",
    archivePrefix = "arXiv",
    primaryClass = "astro-ph.SR",
    doi = "10.3847/2041-8213/ab339b",
    journal = "Astrophys. J. Lett.",
    volume = "881",
    number = "1",
    pages = "L1",
    year = "2019"
}

@article{Bavera:2020inc,
    author = "Bavera, Simone S. and Fragos, Tassos and Qin, Ying and Zapartas, Emmanouil and Neijssel, Coenraad J. and Mandel, Ilya and Batta, Aldo and Gaebel, Sebastian M. and Kimball, Chase and Stevenson, Simon",
    title = "{The origin of spin in binary black holes: Predicting the distributions of the main observables of Advanced LIGO}",
    eprint = "1906.12257",
    archivePrefix = "arXiv",
    primaryClass = "astro-ph.HE",
    doi = "10.1051/0004-6361/201936204",
    journal = "Astron. Astrophys.",
    volume = "635",
    pages = "A97",
    year = "2020"
}

@article{Fuller:2022ysb,
    author = "Fuller, Jim and Lu, Wenbin",
    title = "{The spins of compact objects born from helium stars in binary systems}",
    eprint = "2201.08407",
    archivePrefix = "arXiv",
    primaryClass = "astro-ph.HE",
    doi = "10.1093/mnras/stac317",
    journal = "Mon. Not. Roy. Astron. Soc.",
    volume = "511",
    number = "3",
    pages = "3951--3964",
    year = "2022"
}

@article{Fishbach:2018edt,
    author = "Fishbach, Maya and Holz, Daniel E. and Farr, Will M.",
    title = "{Does the Black Hole Merger Rate Evolve with Redshift?}",
    eprint = "1805.10270",
    archivePrefix = "arXiv",
    primaryClass = "astro-ph.HE",
    doi = "10.3847/2041-8213/aad800",
    journal = "Astrophys. J. Lett.",
    volume = "863",
    number = "2",
    pages = "L41",
    year = "2018"
}

@article{Cutler:1994ys,
    author = "Cutler, Curt and Flanagan, Eanna E.",
    title = "{Gravitational waves from merging compact binaries: How accurately can one extract the binary's parameters from the inspiral wave form?}",
    eprint = "gr-qc/9402014",
    archivePrefix = "arXiv",
    reportNumber = "GRP-369",
    doi = "10.1103/PhysRevD.49.2658",
    journal = "Phys. Rev. D",
    volume = "49",
    pages = "2658--2697",
    year = "1994"
}

@article{Baird:2012cu,
    author = "Baird, Emily and Fairhurst, Stephen and Hannam, Mark and Murphy, Patricia",
    title = "{Degeneracy between mass and spin in black-hole-binary waveforms}",
    eprint = "1211.0546",
    archivePrefix = "arXiv",
    primaryClass = "gr-qc",
    doi = "10.1103/PhysRevD.87.024035",
    journal = "Phys. Rev. D",
    volume = "87",
    number = "2",
    pages = "024035",
    year = "2013"
}

@article{Ng:2018neg,
    author = "Ng, Ken K. Y. and Vitale, Salvatore and Zimmerman, Aaron and Chatziioannou, Katerina and Gerosa, Davide and Haster, Carl-Johan",
    title = "{Gravitational-wave astrophysics with effective-spin measurements: asymmetries and selection biases}",
    eprint = "1805.03046",
    archivePrefix = "arXiv",
    primaryClass = "gr-qc",
    doi = "10.1103/PhysRevD.98.083007",
    journal = "Phys. Rev. D",
    volume = "98",
    number = "8",
    pages = "083007",
    year = "2018"
}

@article{Veitch:2014wba,
    author = "Veitch, J. and others",
    title = "{Parameter estimation for compact binaries with ground-based gravitational-wave observations using the LALInference software library}",
    eprint = "1409.7215",
    archivePrefix = "arXiv",
    primaryClass = "gr-qc",
    reportNumber = "LIGO-P1400152",
    doi = "10.1103/PhysRevD.91.042003",
    journal = "Phys. Rev. D",
    volume = "91",
    number = "4",
    pages = "042003",
    year = "2015"
}

@article{Rinaldi:2021bhm,
    author = "Rinaldi, Stefano and Del Pozzo, Walter",
    title = "{(H)DPGMM: a hierarchy of Dirichlet process Gaussian mixture models for the inference of the black hole mass function}",
    eprint = "2109.05960",
    archivePrefix = "arXiv",
    primaryClass = "astro-ph.IM",
    doi = "10.1093/mnras/stab3224",
    journal = "Mon. Not. Roy. Astron. Soc.",
    volume = "509",
    number = "4",
    pages = "5454--5466",
    year = "2021"
}

@article{Tiwari:2020otp,
    author = "{Tiwari}, Vaibhav and {Fairhurst}, Stephen",
    title = "{The Emergence of Structure in the Binary Black Hole Mass Distribution}",
    eprint = "2011.04502",
    archivePrefix = "arXiv",
    primaryClass = "astro-ph.HE",
    doi = "10.3847/2041-8213/abfbe7",
    journal = "Astrophys. J. Lett.",
    volume = "913",
    number = "2",
    pages = "L19",
    year = "2021"
}

@article{Tiwari:2020vym,
    author = "Tiwari, Vaibhav",
    title = "{VAMANA: modeling binary black hole population with minimal assumptions}",
    eprint = "2006.15047",
    archivePrefix = "arXiv",
    primaryClass = "astro-ph.HE",
    doi = "10.1088/1361-6382/ac0b54",
    journal = "Class. Quant. Grav.",
    volume = "38",
    number = "15",
    pages = "155007",
    year = "2021"
}

@article{Rodriguez:2016vmx,
    author = "Rodriguez, Carl L. and Zevin, Michael and Pankow, Chris and Kalogera, Vasilliki and Rasio, Frederic A.",
    title = "{Illuminating Black Hole Binary Formation Channels with Spins in Advanced LIGO}",
    eprint = "1609.05916",
    archivePrefix = "arXiv",
    primaryClass = "astro-ph.HE",
    doi = "10.3847/2041-8205/832/1/L2",
    journal = "Astrophys. J. Lett.",
    volume = "832",
    number = "1",
    pages = "L2",
    year = "2016"
}

@article{Rodriguez:2016kxx,
    author = "Rodriguez, Carl L. and Chatterjee, Sourav and Rasio, Frederic A.",
    title = "{Binary Black Hole Mergers from Globular Clusters: Masses, Merger Rates, and the Impact of Stellar Evolution}",
    eprint = "1602.02444",
    archivePrefix = "arXiv",
    primaryClass = "astro-ph.HE",
    doi = "10.1103/PhysRevD.93.084029",
    journal = "Phys. Rev. D",
    volume = "93",
    number = "8",
    pages = "084029",
    year = "2016"
}

@article{PortegiesZwart:1999nm,
    author = "Portegies Zwart, Simon F. and McMillan, Stephen",
    title = "{Black hole mergers in the universe}",
    eprint = "astro-ph/9910061",
    archivePrefix = "arXiv",
    doi = "10.1086/312422",
    journal = "Astrophys. J. Lett.",
    volume = "528",
    pages = "L17",
    year = "2000"
}

@article{Yang:2019cbr,
    author = "Yang, Yang and others",
    title = "{Hierarchical Black Hole Mergers in Active Galactic Nuclei}",
    eprint = "1906.09281",
    archivePrefix = "arXiv",
    primaryClass = "astro-ph.HE",
    doi = "10.1103/PhysRevLett.123.181101",
    journal = "Phys. Rev. Lett.",
    volume = "123",
    number = "18",
    pages = "181101",
    year = "2019"
}

@article{Tichy:2008du,
    author = "Tichy, Wolfgang and Marronetti, Pedro",
    title = "{The Final mass and spin of black hole mergers}",
    eprint = "0807.2985",
    archivePrefix = "arXiv",
    primaryClass = "gr-qc",
    doi = "10.1103/PhysRevD.78.081501",
    journal = "Phys. Rev. D",
    volume = "78",
    pages = "081501",
    year = "2008"
}

@article{Gerosa:2017kvu,
    author = "Gerosa, Davide and Berti, Emanuele",
    title = "{Are merging black holes born from stellar collapse or previous mergers?}",
    eprint = "1703.06223",
    archivePrefix = "arXiv",
    primaryClass = "gr-qc",
    doi = "10.1103/PhysRevD.95.124046",
    journal = "Phys. Rev. D",
    volume = "95",
    number = "12",
    pages = "124046",
    year = "2017"
}

@article{Tagawa:2021ofj,
    author = "Tagawa, Hiromichi and Haiman, Zolt{\'a}n and Bartos, Imre and Kocsis, Bence and Omukai, Kazuyuki",
    title = "{Signatures of hierarchical mergers in black hole spin and mass distribution}",
    eprint = "2104.09510",
    archivePrefix = "arXiv",
    primaryClass = "astro-ph.HE",
    doi = "10.1093/mnras/stab2315",
    journal = "Mon. Not. Roy. Astron. Soc.",
    volume = "507",
    number = "3",
    pages = "3362--3380",
    year = "2021"
}

@article{Qin:2018vaa,
    author = "Qin, Y. and Fragos, T. and Meynet, G. and Andrews, J. and S{\o}rensen, M. and Song, H. F.",
    title = "{The spin of the second-born black hole in coalescing binary black holes}",
    eprint = "1802.05738",
    archivePrefix = "arXiv",
    primaryClass = "astro-ph.SR",
    doi = "10.1051/0004-6361/201832839",
    journal = "Astron. Astrophys.",
    volume = "616",
    pages = "A28",
    year = "2018"
}

@article{Antonini:2017tgo,
    author = "Antonini, Fabio and Rodriguez, Carl L. and Petrovich, Cristobal and Fischer, Caitlin L.",
    title = "{Precessional dynamics of black hole triples: binary mergers with near-zero effective spin}",
    eprint = "1711.07142",
    archivePrefix = "arXiv",
    primaryClass = "astro-ph.HE",
    doi = "10.1093/mnrasl/sly126",
    journal = "Mon. Not. Roy. Astron. Soc.",
    volume = "480",
    number = "1",
    pages = "L58--L62",
    year = "2018"
}

@article{Ma:2023nrf,
    author = "Ma, Linhao and Fuller, Jim",
    title = "{Tidal Spin-up of Black Hole Progenitor Stars}",
    eprint = "2305.08356",
    archivePrefix = "arXiv",
    primaryClass = "astro-ph.HE",
    doi = "10.3847/1538-4357/acdb74",
    journal = "Astrophys. J.",
    volume = "952",
    number = "1",
    pages = "53",
    year = "2023",
    note = "[Erratum: Astrophys.J. 965, (2024)]"
}

@dataset{ligo_scientific_collaboration_and_virgo_2021_5546663,
  author       = {{LIGO Scientific, Virgo and KAGRA Collaborations}},
  title        = {{GWTC-3: Parameter estimation data release}},
  month        = nov,
  year         = 2021,
  publisher    = {Zenodo},
  doi          = {10.5281/zenodo.5546663},
  url          = {https://doi.org/10.5281/zenodo.5546663}
}

@dataset{ligo_scientific_collaboration_and_virgo_2023_7890398,
  author       = {{LIGO Scientific, Virgo and KAGRA Collaborations}},
  title        = {{GWTC-3: O1+O2+O3 Search Sensitivity
                   Estimates}},
  month        = nov,
  year         = 2021,
  publisher    = {Zenodo},
  doi          = {10.5281/zenodo.7890398},
  url          = {https://doi.org/10.5281/zenodo.7890398}
}

@dataset{ligo_scientific_collaboration_and_virgo_2023_7890437,
  author       = {{LIGO Scientific, Virgo and KAGRA Collaborations}},
  title        = {GWTC-3: O3 search sensitivity estimates},
  month        = may,
  year         = 2023,
  publisher    = {Zenodo},
  doi          = {10.5281/zenodo.7890437},
  url          = {https://doi.org/10.5281/zenodo.7890437},
}

@dataset{ligo_scientific_collaboration_and_virgo_2021_5117970,
  author       = {{LIGO Scientific and Virgo Collaborations}},
  title        = {{GWTC-2.1: Candidate
                   Data Release}},
  month        = jul,
  year         = 2021,
  publisher    = {Zenodo},
  version      = {v2},
  doi          = {10.5281/zenodo.5117970},
  url          = {https://doi.org/10.5281/zenodo.5117970}
}

@dataset{190521_pe,
  author       = {{LIGO Scientific and Virgo Collaborations}},
  title        = {GW190521 parameter estimation samples and figure data},
  year         = 2020,
  url          = {https://dcc.ligo.org/LIGO-P2000020/public}
}

@article{LIGOScientific:2020iuh,
    author = "Abbott, R. and others",
    collaboration = "LIGO Scientific, Virgo",
    title = "{GW190521: A Binary Black Hole Merger with a Total Mass of $150  M_{\odot}$}",
    eprint = "2009.01075",
    archivePrefix = "arXiv",
    primaryClass = "gr-qc",
    doi = "10.1103/PhysRevLett.125.101102",
    journal = "Phys. Rev. Lett.",
    volume = "125",
    number = "10",
    pages = "101102",
    year = "2020"
}

@article{LIGOScientific:2014pky,
    author = "Aasi, J. and others",
    collaboration = "LIGO Scientific",
    title = "{Advanced LIGO}",
    eprint = "1411.4547",
    archivePrefix = "arXiv",
    primaryClass = "gr-qc",
    doi = "10.1088/0264-9381/32/7/074001",
    journal = "Class. Quant. Grav.",
    volume = "32",
    pages = "074001",
    year = "2015"
}

@article{VIRGO:2014yos,
    author = "Acernese, F. and others",
    collaboration = "VIRGO",
    title = "{Advanced Virgo: a second-generation interferometric gravitational wave detector}",
    eprint = "1408.3978",
    archivePrefix = "arXiv",
    primaryClass = "gr-qc",
    doi = "10.1088/0264-9381/32/2/024001",
    journal = "Class. Quant. Grav.",
    volume = "32",
    number = "2",
    pages = "024001",
    year = "2015"
}

@article{Woosley:2021xba,
    author = "Woosley, S. E. and Heger, Alexander",
    title = "{The Pair-Instability Mass Gap for Black Holes}",
    eprint = "2103.07933",
    archivePrefix = "arXiv",
    primaryClass = "astro-ph.SR",
    doi = "10.3847/2041-8213/abf2c4",
    journal = "Astrophys. J. Lett.",
    volume = "912",
    number = "2",
    pages = "L31",
    year = "2021"
}

@article{Mandel:2009pc,
    author = "Mandel, Ilya",
    title = "{Parameter estimation on gravitational waves from multiple coalescing binaries}",
    eprint = "0912.5531",
    archivePrefix = "arXiv",
    primaryClass = "astro-ph.HE",
    doi = "10.1103/PhysRevD.81.084029",
    journal = "Phys. Rev. D",
    volume = "81",
    pages = "084029",
    year = "2010"
}

@article{Powell:2019nmw,
    author = "Powell, Jade and Stevenson, Simon and Mandel, Ilya and Tino, Peter",
    title = "{Unmodelled Clustering Methods for Gravitational Wave Populations of Compact Binary Mergers}",
    eprint = "1905.04825",
    archivePrefix = "arXiv",
    primaryClass = "astro-ph.HE",
    doi = "10.1093/mnras/stz1938",
    journal = "Mon. Not. Roy. Astron. Soc.",
    volume = "488",
    number = "3",
    pages = "3810--3817",
    year = "2019"
}

@article{Vitale:2015tea,
    author = "Vitale, Salvatore and Lynch, Ryan and Sturani, Riccardo and Graff, Philip",
    title = "{Use of gravitational waves to probe the formation channels of compact binaries}",
    eprint = "1503.04307",
    archivePrefix = "arXiv",
    primaryClass = "gr-qc",
    reportNumber = "LIGO-DOCUMENT-P1500022, LIGO-P1500022",
    doi = "10.1088/1361-6382/aa552e",
    journal = "Class. Quant. Grav.",
    volume = "34",
    number = "3",
    pages = "03LT01",
    year = "2017"
}

@article{Farr:2017gtv,
    author = "Farr, Ben and Holz, Daniel E. and Farr, Will M.",
    title = "{Using Spin to Understand the Formation of LIGO and Virgo's Black Holes}",
    eprint = "1709.07896",
    archivePrefix = "arXiv",
    primaryClass = "astro-ph.HE",
    doi = "10.3847/2041-8213/aaaa64",
    journal = "Astrophys. J. Lett.",
    volume = "854",
    number = "1",
    pages = "L9",
    year = "2018"
}

@article{Farr:2017uvj,
    author = "Farr, Will M. and Stevenson, Simon and Coleman Miller, M. and Mandel, Ilya and Farr, Ben and Vecchio, Alberto",
    title = "{Distinguishing Spin-Aligned and Isotropic Black Hole Populations With Gravitational Waves}",
    eprint = "1706.01385",
    archivePrefix = "arXiv",
    primaryClass = "astro-ph.HE",
    reportNumber = "LIGO-P1700067",
    doi = "10.1038/nature23453",
    journal = "Nature",
    volume = "548",
    pages = "426",
    year = "2017"
}

@article{LIGOScientific:2021usb,
    author = "Abbott, R. and others",
    collaboration = "LIGO Scientific, VIRGO",
    title = "{GWTC-2.1: Deep extended catalog of compact binary coalescences observed by LIGO and Virgo during the first half of the third observing run}",
    eprint = "2108.01045",
    archivePrefix = "arXiv",
    primaryClass = "gr-qc",
    reportNumber = "LIGO-P2100063",
    doi = "10.1103/PhysRevD.109.022001",
    journal = "Phys. Rev. D",
    volume = "109",
    number = "2",
    pages = "022001",
    year = "2024"
}

@article{Tiwari:2021yvr,
    author = "Tiwari, Vaibhav",
    title = "{Exploring Features in the Binary Black Hole Population}",
    eprint = "2111.13991",
    archivePrefix = "arXiv",
    primaryClass = "astro-ph.HE",
    doi = "10.3847/1538-4357/ac589a",
    journal = "Astrophys. J.",
    volume = "928",
    number = "2",
    pages = "155",
    year = "2022"
}

@article{Edelman:2022ydv,
    author = "Edelman, Bruce and Farr, Ben and Doctor, Zoheyr",
    title = "{Cover Your Basis: Comprehensive Data-driven Characterization of the Binary Black Hole Population}",
    eprint = "2210.12834",
    archivePrefix = "arXiv",
    primaryClass = "astro-ph.HE",
    reportNumber = "LIGO-P2200312",
    doi = "10.3847/1538-4357/acb5ed",
    journal = "Astrophys. J.",
    volume = "946",
    number = "1",
    pages = "16",
    year = "2023"
}

@article{Callister:2023tgi,
    author = "Callister, Thomas A. and Farr, Will M.",
    title = "{Parameter-Free Tour of the Binary Black Hole Population}",
    eprint = "2302.07289",
    archivePrefix = "arXiv",
    primaryClass = "astro-ph.HE",
    doi = "10.1103/PhysRevX.14.021005",
    journal = "Phys. Rev. X",
    volume = "14",
    number = "2",
    pages = "021005",
    year = "2024"
}

@article{Toubiana:2023egi,
    author = "Toubiana, Alexandre and Katz, Michael L. and Gair, Jonathan R.",
    title = "{Is there an excess of black holes around 20 M\ensuremath{\odot}? Optimizing the complexity of population models with the use of reversible jump MCMC.}",
    eprint = "2305.08909",
    archivePrefix = "arXiv",
    primaryClass = "gr-qc",
    doi = "10.1093/mnras/stad2215",
    journal = "Mon. Not. Roy. Astron. Soc.",
    volume = "524",
    number = "4",
    pages = "5844--5853",
    year = "2023"
}

@article{Callister:2021fpo,
    author = "Callister, Thomas A. and Haster, Carl-Johan and Ng, Ken K. Y. and Vitale, Salvatore and Farr, Will M.",
    title = "{Who Ordered That? Unequal-mass Binary Black Hole Mergers Have Larger Effective Spins}",
    eprint = "2106.00521",
    archivePrefix = "arXiv",
    primaryClass = "astro-ph.HE",
    doi = "10.3847/2041-8213/ac2ccc",
    journal = "Astrophys. J. Lett.",
    volume = "922",
    number = "1",
    pages = "L5",
    year = "2021"
}

@article{LIGOScientific:2020zkf,
    author = "Abbott, R. and others",
    collaboration = "LIGO Scientific, Virgo",
    title = "{GW190814: Gravitational Waves from the Coalescence of a 23 Solar Mass Black Hole with a 2.6 Solar Mass Compact Object}",
    eprint = "2006.12611",
    archivePrefix = "arXiv",
    primaryClass = "astro-ph.HE",
    reportNumber = "LIGO-P190814",
    doi = "10.3847/2041-8213/ab960f",
    journal = "Astrophys. J. Lett.",
    volume = "896",
    number = "2",
    pages = "L44",
    year = "2020"
}

@article{Thrane:2018qnx,
    author = "Thrane, Eric and Talbot, Colm",
    title = "{An introduction to Bayesian inference in gravitational-wave astronomy: parameter estimation, model selection, and hierarchical models}",
    eprint = "1809.02293",
    archivePrefix = "arXiv",
    primaryClass = "astro-ph.IM",
    doi = "10.1017/pasa.2019.2",
    journal = "Publ. Astron. Soc. Austral.",
    volume = "36",
    pages = "e010",
    year = "2019",
    note = "[Erratum: Publ.Astron.Soc.Austral. 37, e036 (2020)]"
}

@article{vanSon:2021zpk,
    author = "van Son, L. A. C. and de Mink, S. E. and Callister, T. and Justham, S. and Renzo, M. and Wagg, T. and Broekgaarden, F. S. and Kummer, F. and Pakmor, R. and Mandel, I.",
    title = "{The Redshift Evolution of the Binary Black Hole Merger Rate: A Weighty Matter}",
    eprint = "2110.01634",
    archivePrefix = "arXiv",
    primaryClass = "astro-ph.HE",
    doi = "10.3847/1538-4357/ac64a3",
    journal = "Astrophys. J.",
    volume = "931",
    number = "1",
    pages = "17",
    year = "2022"
}

@ARTICLE{2022ApJ...940..184V,
       author = {{van Son}, L.~A.~C. and {de Mink}, S.~E. and {Renzo}, M. and {Justham}, S. and {Zapartas}, E. and {Breivik}, K. and {Callister}, T. and {Farr}, W.~M. and {Conroy}, C.},
        title = "{No Peaks without Valleys: The Stable Mass Transfer Channel for Gravitational-wave Sources in Light of the Neutron Star-Black Hole Mass Gap}",
      journal = {\apj},
         year = 2022,
        month = dec,
       volume = {940},
       number = {2},
          eid = {184},
        pages = {184},
          doi = {10.3847/1538-4357/ac9b0a},
archivePrefix = {arXiv},
       eprint = {2209.13609},
 primaryClass = {astro-ph.HE},
       adsurl = {https://ui.adsabs.harvard.edu/abs/2022ApJ...940..184V}
}

@article{Zevin:2020gbd,
    author = "Zevin, Michael and Bavera, Simone S. and Berry, Christopher P. L. and Kalogera, Vicky and Fragos, Tassos and Marchant, Pablo and Rodriguez, Carl L. and Antonini, Fabio and Holz, Daniel E. and Pankow, Chris",
    title = "{One Channel to Rule Them All? Constraining the Origins of Binary Black Holes Using Multiple Formation Pathways}",
    eprint = "2011.10057",
    archivePrefix = "arXiv",
    primaryClass = "astro-ph.HE",
    doi = "10.3847/1538-4357/abe40e",
    journal = "Astrophys. J.",
    volume = "910",
    number = "2",
    pages = "152",
    year = "2021"
}

@article{Cheng:2023ddt,
    author = "Cheng, April Qiu and Zevin, Michael and Vitale, Salvatore",
    title = "{What You Don\textquoteright{}t Know Can Hurt You: Use and Abuse of Astrophysical Models in Gravitational-wave Population Analyses}",
    eprint = "2307.03129",
    archivePrefix = "arXiv",
    primaryClass = "astro-ph.HE",
    reportNumber = "LIGO DCC: P2300200",
    doi = "10.3847/1538-4357/aced98",
    journal = "Astrophys. J.",
    volume = "955",
    number = "2",
    pages = "127",
    year = "2023"
}

@article{Sigurdsson1993,
  author       = {Sigurdsson, Steinn and Hernquist, Lars},
  title        = {Primordial black holes in globular clusters},
  journal      = {Nature},
  year         = {1993},
  month        = {July},
  volume       = {364},
  number       = {6436},
  pages        = {423--425},
  doi          = {10.1038/364423a0},
  url          = {https://ui.adsabs.harvard.edu/abs/1993Natur.364..423S},
  note         = {Provided by the SAO/NASA Astrophysics Data System}
}

@article{Silsbee:2016djf,
    author = "Silsbee, Kedron and Tremaine, Scott",
    title = "{Lidov-Kozai Cycles with Gravitational Radiation: Merging Black Holes in Isolated Triple Systems}",
    eprint = "1608.07642",
    archivePrefix = "arXiv",
    primaryClass = "astro-ph.HE",
    doi = "10.3847/1538-4357/aa5729",
    journal = "Astrophys. J.",
    volume = "836",
    number = "1",
    pages = "39",
    year = "2017"
}

@article{Stone:2016wzz,
    author = "Stone, Nicholas C. and Metzger, Brian D. and Haiman, Zolt{\'a}n",
    title = "{Assisted inspirals of stellar mass black holes embedded in AGN discs: solving the {\textquoteleft}final au problem{\textquoteright}}",
    eprint = "1602.04226",
    archivePrefix = "arXiv",
    primaryClass = "astro-ph.GA",
    doi = "10.1093/mnras/stw2260",
    journal = "Mon. Not. Roy. Astron. Soc.",
    volume = "464",
    number = "1",
    pages = "946--954",
    year = "2017"
}

@article{Baibhav:2022qxm,
    author = "Baibhav, Vishal and Doctor, Zoheyr and Kalogera, Vicky",
    title = "{Dropping Anchor: Understanding the Populations of Binary Black Holes with Random and Aligned-spin Orientations}",
    eprint = "2212.12113",
    archivePrefix = "arXiv",
    primaryClass = "astro-ph.HE",
    doi = "10.3847/1538-4357/acbf4c",
    journal = "Astrophys. J.",
    volume = "946",
    number = "1",
    pages = "50",
    year = "2023"
}

@article{Mapelli:2020vfa,
    author = "Mapelli, Michela",
    title = "{Binary Black Hole Mergers: Formation and Populations}",
    eprint = "2105.12455",
    archivePrefix = "arXiv",
    primaryClass = "astro-ph.HE",
    doi = "10.3389/fspas.2020.00038",
    journal = "Front. Astron. Space Sci.",
    volume = "7",
    pages = "38",
    year = "2020"
}

@article{Baibhav:2020xdf,
    author = "Baibhav, Vishal and Gerosa, Davide and Berti, Emanuele and Wong, Kaze W. K. and Helfer, Thomas and Mould, Matthew",
    title = "{The mass gap, the spin gap, and the origin of merging binary black holes}",
    eprint = "2004.00650",
    archivePrefix = "arXiv",
    primaryClass = "astro-ph.HE",
    doi = "10.1103/PhysRevD.102.043002",
    journal = "Phys. Rev. D",
    volume = "102",
    number = "4",
    pages = "043002",
    year = "2020"
}

@article{Biscoveanu:2022qac,
    author = "Biscoveanu, Sylvia and Callister, Thomas A. and Haster, Carl-Johan and Ng, Ken K. Y. and Vitale, Salvatore and Farr, Will M.",
    title = "{The Binary Black Hole Spin Distribution Likely Broadens with Redshift}",
    eprint = "2204.01578",
    archivePrefix = "arXiv",
    primaryClass = "astro-ph.HE",
    reportNumber = "LIGO document number LIGO-P2200105",
    doi = "10.3847/2041-8213/ac71a8",
    journal = "Astrophys. J. Lett.",
    volume = "932",
    number = "2",
    pages = "L19",
    year = "2022"
}

@article{Bavera:2022mef,
    author = "Bavera, Simone S. and Fishbach, Maya and Zevin, Michael and Zapartas, Emmanouil and Fragos, Tassos",
    title = "{The \ensuremath{\chi}eff \ensuremath{-} z correlation of field binary black hole mergers and how 3G gravitational-wave detectors can constrain it}",
    eprint = "2204.02619",
    archivePrefix = "arXiv",
    primaryClass = "astro-ph.HE",
    doi = "10.1051/0004-6361/202243724",
    journal = "Astron. Astrophys.",
    volume = "665",
    pages = "A59",
    year = "2022"
}

@article{Franciolini:2022iaa,
    author = "Franciolini, Gabriele and Pani, Paolo",
    title = "{Searching for mass-spin correlations in the population of gravitational-wave events: The GWTC-3 case study}",
    eprint = "2201.13098",
    archivePrefix = "arXiv",
    primaryClass = "astro-ph.HE",
    doi = "10.1103/PhysRevD.105.123024",
    journal = "Phys. Rev. D",
    volume = "105",
    number = "12",
    pages = "123024",
    year = "2022"
}

@article{Li:2023yyt,
    author = "Li, Yin-Jie and Wang, Yuan-Zhu and Tang, Shao-Peng and Fan, Yi-Zhong",
    title = "{Resolving the Stellar-Collapse and Hierarchical-Merger Origins of the Coalescing Black Holes}",
    eprint = "2303.02973",
    archivePrefix = "arXiv",
    primaryClass = "astro-ph.HE",
    doi = "10.1103/PhysRevLett.133.051401",
    journal = "Phys. Rev. Lett.",
    volume = "133",
    number = "5",
    pages = "051401",
    year = "2024"
}

@article{Li:2024jzi,
    author = "Li, Yin-Jie and Tang, Shao-Peng and Gao, Shi-Jie and Wu, Dao-Cheng and Wang, Yuan-Zhu",
    title = "{Exploring Field-evolution and Dynamical-capture Coalescing Binary Black Holes in GWTC-3}",
    eprint = "2404.09668",
    archivePrefix = "arXiv",
    primaryClass = "astro-ph.HE",
    doi = "10.3847/1538-4357/ad83b5",
    journal = "Astrophys. J.",
    volume = "977",
    number = "1",
    pages = "67",
    year = "2024"
}

@article{Pierra:2024fbl,
    author = "Pierra, Gr\'egoire and Mastrogiovanni, Simone and Perri\`es, St\'ephane",
    title = "{The spin magnitude of stellar-mass black holes evolves with the mass}",
    eprint = "2406.01679",
    archivePrefix = "arXiv",
    primaryClass = "gr-qc",
    doi = "10.1051/0004-6361/202452545",
    journal = "Astron. Astrophys.",
    volume = "692",
    pages = "A80",
    year = "2024"
}

@article{Adamcewicz:2023mov,
    author = "Adamcewicz, Christian and Lasky, Paul D. and Thrane, Eric",
    title = "{Evidence for a Correlation between Binary Black Hole Mass Ratio and Black Hole Spins}",
    eprint = "2307.15278",
    archivePrefix = "arXiv",
    primaryClass = "astro-ph.HE",
    doi = "10.3847/1538-4357/acf763",
    journal = "Astrophys. J.",
    volume = "958",
    number = "1",
    pages = "13",
    year = "2023"
}

@article{Payne:2022xan,
    author = "Payne, Ethan and Thrane, Eric",
    title = "{Model exploration in gravitational-wave astronomy with the maximum population likelihood}",
    eprint = "2210.11641",
    archivePrefix = "arXiv",
    primaryClass = "astro-ph.IM",
    doi = "10.1103/PhysRevResearch.5.023013",
    journal = "Phys. Rev. Res.",
    volume = "5",
    number = "2",
    pages = "023013",
    year = "2023"
}

@article{Payne:2024ywe,
    author = "Payne, Ethan and Kremer, Kyle and Zevin, Michael",
    title = "{Spin Doctors: How to Diagnose a Hierarchical Merger Origin}",
    eprint = "2402.15066",
    archivePrefix = "arXiv",
    primaryClass = "gr-qc",
    reportNumber = "LIGO DCC P2400050",
    doi = "10.3847/2041-8213/ad3e82",
    journal = "Astrophys. J. Lett.",
    volume = "966",
    number = "1",
    pages = "L16",
    year = "2024"
}

@article{Antonini:2022vib,
    author = "Antonini, Fabio and Gieles, Mark and Dosopoulou, Fani and Chattopadhyay, Debatri",
    title = "{Coalescing black hole binaries from globular clusters: mass distributions and comparison to gravitational wave data from GWTC-3}",
    eprint = "2208.01081",
    archivePrefix = "arXiv",
    primaryClass = "astro-ph.HE",
    doi = "10.1093/mnras/stad972",
    journal = "Mon. Not. Roy. Astron. Soc.",
    volume = "522",
    number = "1",
    pages = "466--476",
    year = "2023"
}

@article{Antonini:2024het,
    author = "Antonini, Fabio and Romero-Shaw, Isobel M. and Callister, Thomas",
    title = "{Star Cluster Population of High Mass Black Hole Mergers in Gravitational Wave Data}",
    eprint = "2406.19044",
    archivePrefix = "arXiv",
    primaryClass = "astro-ph.HE",
    doi = "10.1103/PhysRevLett.134.011401",
    journal = "Phys. Rev. Lett.",
    volume = "134",
    number = "1",
    pages = "011401",
    year = "2025"
}

@article{Godfrey:2023oxb,
      title={Cosmic Cousins: Identification of a Subpopulation of Binary Black Holes Consistent with Isolated Binary Evolution}, 
      author={Jaxen Godfrey and Bruce Edelman and Ben Farr},
      journal={},
      year={2024},
      eprint={2304.01288},
      archivePrefix={arXiv},
      primaryClass={astro-ph.HE},
      url={https://arxiv.org/abs/2304.01288}, 
}

@article{Ray:2024hos,
      title={Searching for binary black hole sub-populations in gravitational wave data using binned Gaussian processes}, 
      author={Anarya Ray and Ignacio Magaña Hernandez and Katelyn Breivik and Jolien Creighton},
      year={2024},
      journal={},
      eprint={2404.03166},
      archivePrefix={arXiv},
      primaryClass={astro-ph.HE},
      url={https://arxiv.org/abs/2404.03166}, 
}

@article{Heinzel:2024jlc,
    author = "Heinzel, Jack and Mould, Matthew and \'Alvarez-L\'opez, Sof\'\i{}a and Vitale, Salvatore",
    title = "{High resolution nonparametric inference of gravitational-wave populations in multiple dimensions}",
    eprint = "2406.16813",
    archivePrefix = "arXiv",
    primaryClass = "astro-ph.HE",
    doi = "10.1103/PhysRevD.111.063043",
    journal = "Phys. Rev. D",
    volume = "111",
    number = "6",
    pages = "063043",
    year = "2025"
}

@article{Heinzel:2024hva,
    author = "Heinzel, Jack and Mould, Matthew and Vitale, Salvatore",
    title = "{Nonparametric analysis of correlations in the binary black hole population with LIGO-Virgo-KAGRA data}",
    eprint = "2406.16844",
    archivePrefix = "arXiv",
    primaryClass = "astro-ph.HE",
    doi = "10.1103/PhysRevD.111.L061305",
    journal = "Phys. Rev. D",
    volume = "111",
    number = "6",
    pages = "L061305",
    year = "2025"
}

@article{Callister:2021gxf,
      title={A Thesaurus for Common Priors in Gravitational-Wave Astronomy}, 
      author={T. A. Callister},
      year={2021},
      journal={},
      eprint={2104.09508},
      archivePrefix={arXiv},
      primaryClass={gr-qc},
      url={https://arxiv.org/abs/2104.09508}, 
}

@article{Mould:2022xeu,
    author = "Mould, Matthew and Gerosa, Davide and Broekgaarden, Floor S. and Steinle, Nathan",
    title = "{Which black hole formed first? Mass-ratio reversal in massive binary stars from gravitational-wave data}",
    eprint = "2205.12329",
    archivePrefix = "arXiv",
    primaryClass = "astro-ph.HE",
    doi = "10.1093/mnras/stac2859",
    journal = "Mon. Not. Roy. Astron. Soc.",
    volume = "517",
    number = "2",
    pages = "2738--2745",
    year = "2022"
}

@article{Olejak:2024qxr,
    author = "Olejak, Aleksandra and Klencki, Jakub and Xu, Xiao-Tian and Wang, Chen and Belczynski, Krzysztof and Lasota, Jean-Pierre",
    title = "{Unequal-mass, highly-spinning binary black hole mergers in the stable mass transfer formation channel}",
    eprint = "2404.12426",
    archivePrefix = "arXiv",
    primaryClass = "astro-ph.HE",
    doi = "10.1051/0004-6361/202450480",
    journal = "Astron. Astrophys.",
    volume = "689",
    pages = "A305",
    year = "2024"
}

@article{Banerjee:2024wbq,
    author = "Banerjee, Sambaran and Olejak, Aleksandra",
    title = "{On the effective spin-mass ratio '$\chi_{\rm eff}-q$' relation of binary black hole mergers that evolved in isolation}",
    journal = {},
    eprint = "2411.15112",
    archivePrefix = "arXiv",
    primaryClass = "astro-ph.HE",
    month = "11",
    year = "2024"
}

@article{Galaudage:2024meo,
    author = "Galaudage, Shanika and Lamberts, Astrid",
    title = "{Compactness peaks: An astrophysical interpretation of the mass distribution of merging binary black holes}",
    eprint = "2407.17561",
    archivePrefix = "arXiv",
    primaryClass = "astro-ph.HE",
    doi = "10.1051/0004-6361/202451654",
    journal = "Astron. Astrophys.",
    volume = "694",
    pages = "A186",
    year = "2025"
}

@article{Fabbri:2025faf,
    author = "Fabbri, Cecilia Maria and Gerosa, Davide and Santini, Alessandro and Mould, Matthew and Toubiana, Alexandre and Gair, Jonathan",
    title = "{Reconstructing parametric gravitational-wave population fits from non-parametric results without refitting the data}",
    eprint = "2501.17233",
    archivePrefix = "arXiv",
    primaryClass = "astro-ph.HE",
    journal = "",
    month = "1",
    year = "2025"
}


\end{document}